\documentclass[referee]{raa}
\usepackage{threeparttable}
\usepackage{multirow}
\usepackage[figuresright]{rotating}
\usepackage{subfigure}
\usepackage{epic,eepic}
\usepackage{graphicx}
\usepackage{natbib}
\begin{document}

\title{X-RAY ACTIVITY FROM DIFFERENT TYPES OF STARS}
\volnopage{Vol.0 (200x) No.0, 000--000}      %%preserved for Editor. DOn't remove!
\setcounter{page}{1}          %%starting page, preserved for Editor. DOn't remove!

\author{Lin He\inst{1,2}, Song Wang\inst{2}, Xiaojie Xu\inst{1}, Roberto Soria\inst{3,4,5}, Jifeng Liu\inst{2,3,6}, Xiangdong Li\inst{1}, Yu Bai\inst{2}, Zhongrui Bai\inst{2}, Jincheng Guo\inst{7,}\thanks{LAMOST Fellow}, Yanli Qiu\inst{2,3}, Yong Zhang\inst{8},
Ruochuan Xu\inst{9}, \and Kecheng Qian\inst{10}}

\institute{School of Astronomy and Space Science and Key Laboratory of Modern Astronomy and Astrophysics, Nanjing University, Nanjing, 210093, P. R. China
\and
Key Laboratory of Optical Astronomy, National Astronomical Observatories,
Chinese Academy of Sciences, Beijing 100012, China; songw@bao.ac.cn
\and
College of Astronomy and Space Sciences, University of Chinese Academy of Sciences, Beijing 100049, China
\and
International Centre for Radio Astronomy Research, Curtin University, GPO Box U1987, Perth, WA 6845, Australia
\and
Sydney Institute for Astronomy, School of Physics A28, The University of Sydney, Sydney, NSW 2006, Australia
\and
WHU-NAOC Joint Center for Astronomy, Wuhan University, Wuhan, Hubei 430072, China
\and
Department of Astronomy, Peking University, Beijing 100871, China
\and
Nanjing Institute of Astronomical Optics \& Technology, National Astronomical Observatories, Chinese Academy of Sciences, Nanjing 210042, China
\and
International Department, the Affiliated High School of SCNU. No. 1 Zhongshandadao West, Tianhe District, Guangzhou 510630, China
\and
United World College Changshu China. No.88 Kun-Cheng-Hu-Xi Road, Changshu, Jiangsu 215500, China 
}

\abstract
{X-ray emission is an important indicator of stellar activity.
In this paper, we study stellar X-ray activity using the {\it XMM-Newton} and LAMOST data for different types of stars.
We provide a sample including 1259 X-ray emitting stars, of which 1090 have accurate stellar parameter estimations. Our sample size is much larger than those in previous works.
We find a bimodal distribution of X-ray to optical flux ratio (log($f_X/f_V$)) for G and K stars.
We interpret that this bimodality is due to two subpopulations with different coronal heating rates.
Furthermore, using the full widths at half maxima calculated from H$\alpha$ and H$\beta$ lines,
we show that these stars in the inactive peaks have smaller rotational velocities.
This is consistent with
the magnetic dynamo theory that stars with low rotational velocities have low levels of stellar activity.
We also examine the correlation between log($f_X/f_V$) and luminosity of the excess emission in the H$\alpha$ line,
and find a tight relation between the coronal and chromospheric activity indicators.}

\keywords{X-rays: stars -- stars: activity -- stars: late-type}
%\keywords{surveys, X-ray --- stars: stellar activity}

\authorrunning{He et al.}
\titlerunning{STELLAR X-RAY EMISSION}
\maketitle

\section{INTRODUCTION}
\label{intro.sec}

%It has been known from the 1970s that almost all stellar classes are X-ray emitters
Stars of almost all stellar classes are X-ray emitters
\citep{Harnden1979,Stocke1983,Schmitt1995,Rogel2006}.
X-ray emission from late-type main-sequence stars comes from a magnetic corona which contains a plasma at temperature exceeding $\sim$1 million K.
The coronal magnetic structures and heating mechanisms are controlled by surface magnetic fields \citep{2004A&ARv..12...71G}, the latter of which are generally thought caused by a complex dynamo mechanism \citep[e.g.,][]{Pizzolato2003}.

The magnetic dynamo mechanism has been observationally evidenced by
the famous activity-rotation correlation \citep{Skumanich1972}.
\cite{1981ApJ...245..671W} discovered the relation between X-ray luminosity ($L_X$) and rotation for RS CVn systems.
A more accurate relation,
%$L_X$ (ergs s$^{-1}$) $\sim$ 10$^{27}$ ($Vsini$ (km s$^{-1}$))$^{2}$,
$L_X$ $\sim$ 10$^{27}$($Vsini$)$^{2}$,
was given by \citet{1981ApJ...248..279P} for late-type stars.
Furthermore, it was found that the activity-rotation correlation
depends on the stellar mass \citep{Pizzolato2003}.
This is explained as that the generation of magnetic energy by large-scale dynamo action
is driven by rotation and convection \citep[e.g.,][]{Charbonneau2010, Reiners2014}.
On the other hand, there is a saturated X-ray luminosity ($L_X/L_{\rm bol} \approx 10^{-3}$, where $L_{\rm bol}$ is the bolometric luminosity)
for most active stars where $L_X/L_{\rm bol}$ does not change below a certain rotation period \citep{Vilhu1984,1987ApJ...321..958V}.
Two scenarios are often used to explain the saturation and super-saturation \citep[i.e., the activity starts decreasing as stellar rotation rate increases to a critical value;][]{Prosser1996} of stellar activity: polar up-drift migration \citep{Solanki1997} and centrifugal stripping \citep{Jardine1999}.

Although there have been many studies about stellar X-ray emission, some issues are still poorly understood,
such as coronal heating and the evolution of stellar activity.
One main limitation of previous studies is the small sample size of X-ray emitting stars with accurate stellar parameter estimations.
This paper uses the largest spectral database, from observations by the Large Sky Area Multi-object Fiber Spectroscopic Telescope (LAMOST, also named the Guoshoujing Telescope), to present stellar parameters (e.g., effective temperature, surface gravity, metallicity)
for more than 1200 X-ray emitting stars observed by {\it XMM-Newton}.
We will study stellar X-ray activities over a wide range of stellar parameters.
This may help us improve the understanding of these open issues \citep{Testa2015}.

The paper is organized as follows.
In Section \ref{data.sec}, we show the data analysis and sample selection from the 3XMM-DR5 \citep{Rosen2016} and the LAMOST DR3 \citep{Luo2015}.
%In Section \ref{result.sec}, we calculate the X-ray to optical flux ratio, and present catalogs including the X-ray and LAMOST information.
In Section \ref{result.sec}, we calculate the X-ray to optical flux ratio, and study the correlation between the X-ray to optical flux ratio and different stellar parameters.
In Section \ref{bimodality.sec}, we discuss and explain the bimodality of stellar X-ray activity.
Finally, we summarize the results in Section \ref{summary.sec}.

\section{SAMPLE SELECTION AND DATA REDUCTION}
\label{data.sec}

\subsection{Sample Selection}

We cross-matched the 3XMM-DR5 catalog and the LAMOST DR3 catalog using a radius of 3$^{\prime\prime}$.
This led to 3579 unique {\it XMM-Newton} sources with LAMOST spectral observations.
To calculate the likelihood of mismatch,
we shifted the positions of {\it XMM-Newton} sources by 1$^{\prime}$,
and cross-matched them with the LAMOST catalog again using the same radius.
In this case, we obtained 135 matches,
and we conclude the likelihood of mismatch is about 3.77$\%$.

We used several criteria to get a clean sample.
Firstly, for the {\it XMM-Newton} data,
we selected sources with {\sc sum\_flag} $\leq$ 2 and {\sc sc\_extent} $=$ 0.
The former is the summary flag derived from the EPIC warning flags, which is used to exclude spurious detections; the latter is the total band extent that is used to recognize point sources.
%the summary flag of the detection from the EPIC flag was set to be {\sc sum\_flag}  $\leq$ 2, and the total band extent of the source was set to be {\sc sc\_extent} $=$ 0,
%in order to obtain X-ray point sources.
Secondly, for the LAMOST spectra, we only used those with
signal-to-noise ratio (SNR) higher than 7.5 in the $r$ band.
%high signal-to-noise ratios (${\rm snr\_r} \geq 7.5$).
Thirdly, there are four main kinds of classes in the LAMOST database: ``STAR'', ``GALAXY'', ``QSO'', and ``Unknown''.
Spectra flagged as ``Unknown'' were excluded from the sample.
Some other sources, like double stars and white dwarfs, were also excluded according to the classification of the LAMOST catalog.
This led to a sample of 1564 sources, including 134 Galaxies, 60 QSOs, and 1370 stars.
%In order to increase the purity of the single star sample,
Finally, we cross-matched the 1370 stars with the SIMBAD database using a radius of 3 $^{\prime\prime}$.
%After checking and analysing their information given by SIMBAD, we identified
About 100 sources are actually not main-sequence stars:
59 multiple objects, 32 pre-main-sequence stars, 18 globular clusters, one galaxy, and one possible active galactic nucleus (AGN).
All these sources were excluded from the stellar sample.
%We also visually checked these spectra in order to be sure that they were classified correctly.
We visually checked all the spectra of the sample sources (e.g., stars, galaxies, and QSOs), and
the final stellar sample contains 1259 stars.

\subsection{LAMOST data}

LAMOST is a  reflecting Schmidt telescope with a clean aperture of 4 meters and a field of view of 5 degrees \citep{Cui2012, Zhao2012}.
With 4000 fibers, it started its optical spectroscopic survey in 2012, and has successfully accomplished the fourth year mission \citep{2012RAA....12..735D,Luo2015}. The third data release, DR3, contains 5,755,126 spectra, including 5,268,687 stellar spectra, 61,815 galaxy spectra, 16,351 spectra of quasars, and 408,273 spectra of unknown objects \citep{Luo2015}.
In this work, we obtained the effective temperature ($T_{\rm eff}$), metallicity ([Fe/H]), and surface gravity (log($g$)) from the stellar parameter catalogs for A-, F-, G-, and K-type stars. The typical uncertainties of the $T_{\rm eff}$ and log($g$) are about 150 K and 0.3 dex \citep{Wu2011}.
We collected the extinction estimations from \cite{Xiang2017}; the typical uncertainty of $E(B-V)$ in their catalog is about 0.03.
Finally, we calculated the equivalent width (EW) of the H$\alpha$ line,
and the full width at half maximum (FWHM) of H$\alpha$  and H$\beta$ lines.
The EW is calculated using the following formula:
\begin{equation}
{\rm EW} = \int \frac{f(\lambda)-f(0)}{f(0)} d\lambda,
\label{ew.eq}
\end{equation}
where $f(0)$ denotes the nearby pseudo-continuum flux.

\subsection{{\it XMM-Newton} data}

The 3XMM-DR5 catalog \citep{Rosen2016} contains 565,962 X-ray detections comprising 396,910 unique X-ray sources,
which is one of the largest X-ray source catalogues ever produced.
We used PIMMS \footnote{https://heasarc.gsfc.nasa.gov/docs/software/tools/pimms.html}
to convert the 0.2--12 keV count rate (CR) to the unabsorbed 0.3--3.5 keV flux for the PN, M1, and M2, respectively.
For stars we assumed an APEC model with individual absorptions, solar abundance, and a moderate coronal temperature (log$T = $6.5) for the stars \citep{Schmitt1990}.
The individual absorptions for stars were converted form their extinctions \citep{Foight2016},
\begin{equation}
N_H = (2.87\pm 0.12)\times10^{21}A_V\quad {\rm cm^{-2}}.
\end{equation}
The uncertainty of $N_H$ is about 2.67$\times$10$^{20}$ cm$^{-2}$.
We should note that the unabsorbed X-ray flux $f_X$, converted from the count rate using PIMMS,
is dependent on the coronal temperature set in the APEC model.
However, the exact plasma temperature is not accurately known.
To evaluate the influence on $f_X$, we re-estimated it with a higher temperature (log$T = 7$).
Using the mean value of $N_{\rm H}$ ($\approx 10^{21}$ cm$^{-2}$) for the sample stars,
we find $f_X$ decreases by a factor of $\approx$ 0.05.
Therefore, the influence of plasma temperature can be ignored in our study.
For galaxies and QSOs we assumed a power-law model with $\Gamma$ = 1.7 and Galactic foreground absorptions \citep{2011ApJ...737..103S}.
Then, we obtained $f_X$ as the mean value of the three cameras weighted by the errors.
%
%we calculated the X-ray flux (see Section \ref{ratio.sec}) using the 0.2--12 keV count rate for
%the PN, M1, and M2 cameras, respectively, and obtained a mean flux for each star.
We used the {\sc EP\_HR2} as the hardness ratio ($HR$), which is
%defined as the ratio between the count rates of 0.5--1 keV and 1--2 keV
defined as (${\rm CR_{1-2~keV}} - {\rm CR_{0.5-1~keV}}$) $/$ (${\rm CR_{1-2~keV}} + {\rm CR_{0.5-1~keV}}$)
and calculated by averaging over all three cameras.

\section{RESULTS}
\label{result.sec}

\subsection{X-ray to optical flux Ratio}
\label{ratio.sec}

The ranges of X-ray to optical flux ratio are distinctly different for each stellar type, AGNs,  BL Lac objects, clusters of galaxies, and normal galaxies \citep[e.g.,][]{1991ApJS...76..813S,2003AJ....126..575H,2003A&A...406..535Z,2007ApJS..172..353B,2008ApJS..179...19L}.
Using the definition from \citet{Maccacaro1988}, we estimated log($f_X/f_V$) as
\begin{equation}
\label{fx2fo.eq}
\log(f_X/f_V) = \log(f_X) + 0.4V_0 + 5.37,
\end{equation}
where $f_X$ is the unabsorbed 0.3--3.5 keV flux, and $V_0$ is the extinction-corrected $V$-band magnitude.

In order to obtain the $V$-band magnitude, we cross-matched the LAMOST DR3 catalog and the UCAC4 catalog \citep{Zacharias2013} with a radius of 3$^{\prime\prime}$.
For objects without a UCAC4 $V$-band magnitude, we calculated it using the $g$ and $r$ magnitudes from the Sloan Digital Sky Survey (SDSS) following \citet{Jester2005}:
\begin{equation}
V_0 = g_0 - 0.59\times(g-r)_0 - 0.01.
\end{equation}

The errors of log($f_X/f_V$) were calculated as a combination of the errors of X-ray flux, $V$-band magnitude, and extinction.
The X-ray information and the stellar parameters from the LAMOST catalog are summarized in Tables 1 and 2.

\begin{figure*}[!htb]
%\figurenum{2}
\centering
\includegraphics[width=1.05\textwidth]{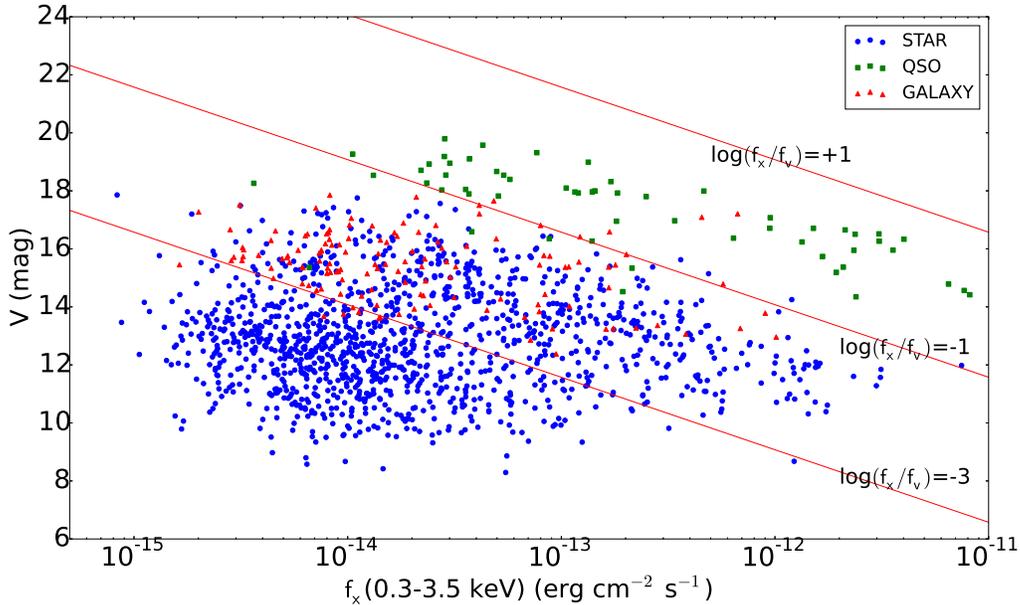}
\caption{$V$ magnitude against $f_X$ (0.3--3.5 keV).
The blue, green, and red points indicate stars, QSOs, and galaxies, which are classified by the LAMOST catalog.
The red lines indicate constant X-ray to optical flux ratios as $+$1, $-$1 and $-$3.}
\label{total_scatter.fig}
\end{figure*}

\subsection{log($f_X/f_V$) for Stellar and Non-stellar Objects}
\label{fx2fv.sec}

Different kinds of objects have their typical ranges of log($f_X/f_V$) values (Figure \ref{total_scatter.fig}).
Most of the stars have log($f_X/f_V$) less than $-$1;
most of the galaxies have log($f_X/f_V$) between $-$3 and $-$1;
most of the QSOs have log($f_X/f_V$) between $-$1 and $+$1.
These ranges are consistent with previous works
\citep[e.g.,][]{1991ApJS...76..813S,1999A&A...350..743K,2004MNRAS.349..135G,2009ApJS..181..444A}.

\begin{figure*}[!htb]
%\figurenum{3}
\center
\includegraphics[width=1.05\textwidth]{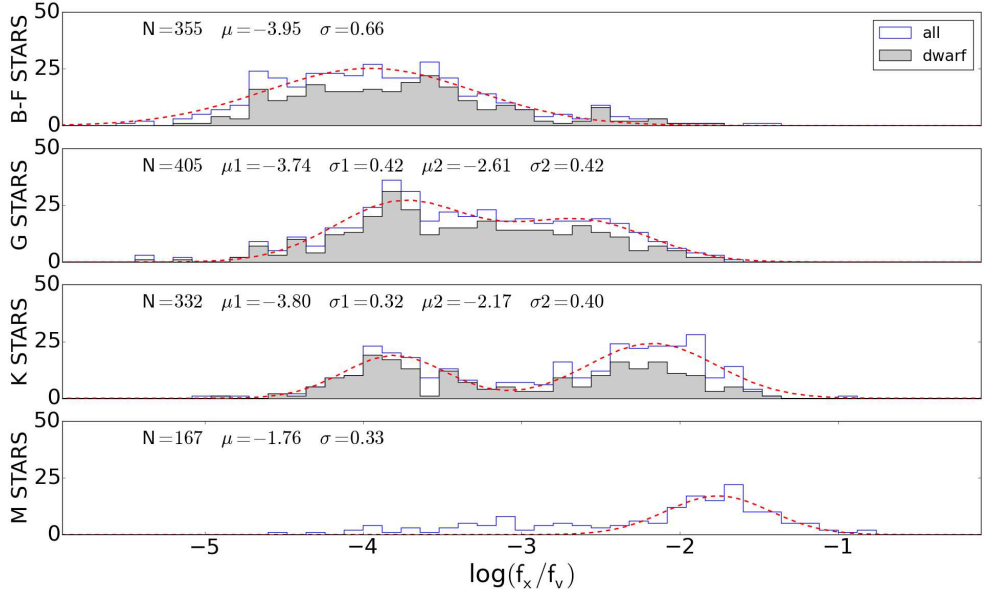}
\caption{log($f_X/f_V$) distributions for B, A, F, G, K and M stars.
The B, A and F groups are shown in one panel due to the small number of stars.
The red dashed lines are the Gaussian fittings to the histograms.
For the B-F and M groups, single-gaussian functions are used for the fitting,
while for the G and K groups, double-gaussian functions are used to fit the log($f_X/f_V$) distributions.
%the results that the distributions are fitted with Gaussian Fitting.
The $N, \mu, \sigma$ indicate the  number of sources, the mean value, and the standard deviation.
Gray histograms represent the dwarfs in each spectral type. For B and M stars, no log($g$) value was given by the LAMOST catalog, thus no giant and dwarf classification was done for them.
%Because of lacking sources of B and A type stars,
}
\label{double_hist.fig}
\end{figure*}

\begin{figure}[htb]
%\figurenum{6}
\centering
\includegraphics[width=1.05\textwidth]{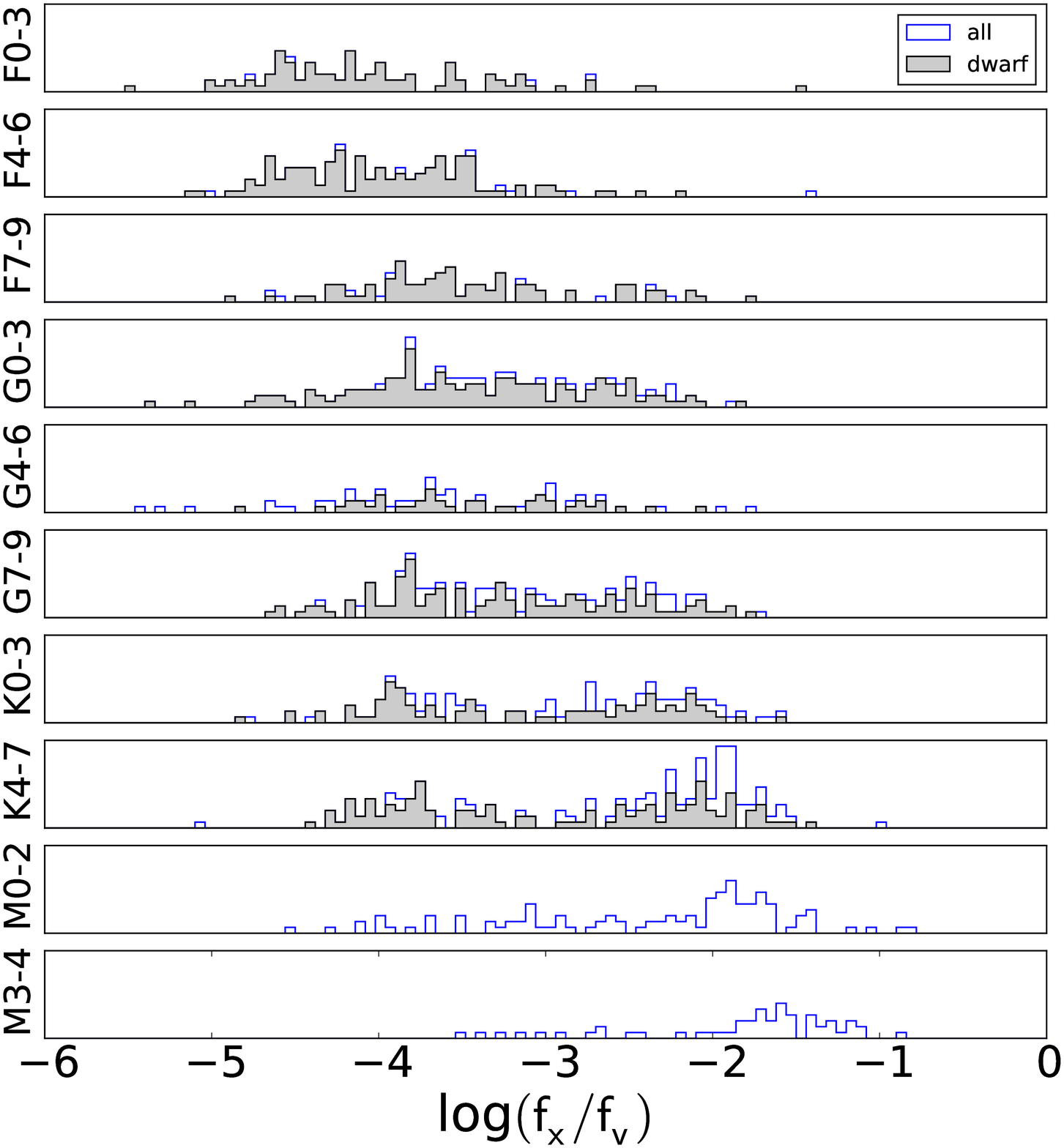}
\caption{log($f_X/f_V$) distributions for subtypes of the F, G, K, and M stars. Gray histograms represent the dwarfs in each spectral type. For M stars, no log($g$) value was given by the LAMOST catalog, thus no giant and dwarf classification was done for them.}
\label{new_narrow.fig}
\end{figure}

For the stellar sample, we divided these stars into four groups according to their spectral types.
Each stellar type shows a widest range of emission levels,
with log($f_X/f_V$) ranging from $\approx~-$5 to $\approx~-$1 (Figure \ref{double_hist.fig}).
Generally, late-type stars have higher log($f_X/f_V$) than early-type stars,
because the optical luminosity decreases more rapidly than the X-ray luminosity for decreasing stellar masses.
The distributions of G and K stars show bimodality,
which is consistent with previous studies \citep{1991ApJS...76..813S,2009ApJS..181..444A,2012ApJ...756...27L}.
For G stars, there are more inactive stars than active ones,
while for K stars, more active stars are apparent.
The log($f_X/f_V$) distributions of K subtypes show clear bimodality (Figure \ref{new_narrow.fig}).
However, due to the sample limit, it is difficult to claim whether the G subtypes show bimodal distributions or broad distributions with local peaks. 
A clear evolutionary trend of the X-ray activity can be seen: from a single inactive distribution (F type), to a weak bimodal distribution (G type), to a clear bimodal distribution (K type), to a single active distribution (M type). Future work with a larger sample may shed more light on the distributions of the subtypes.

\subsection{Comparison with Previous Works}
%\label{xtime.sec}

\begin{figure*}[!htb]
%\figurenum{4}
\center
\includegraphics[width=1.05\textwidth]{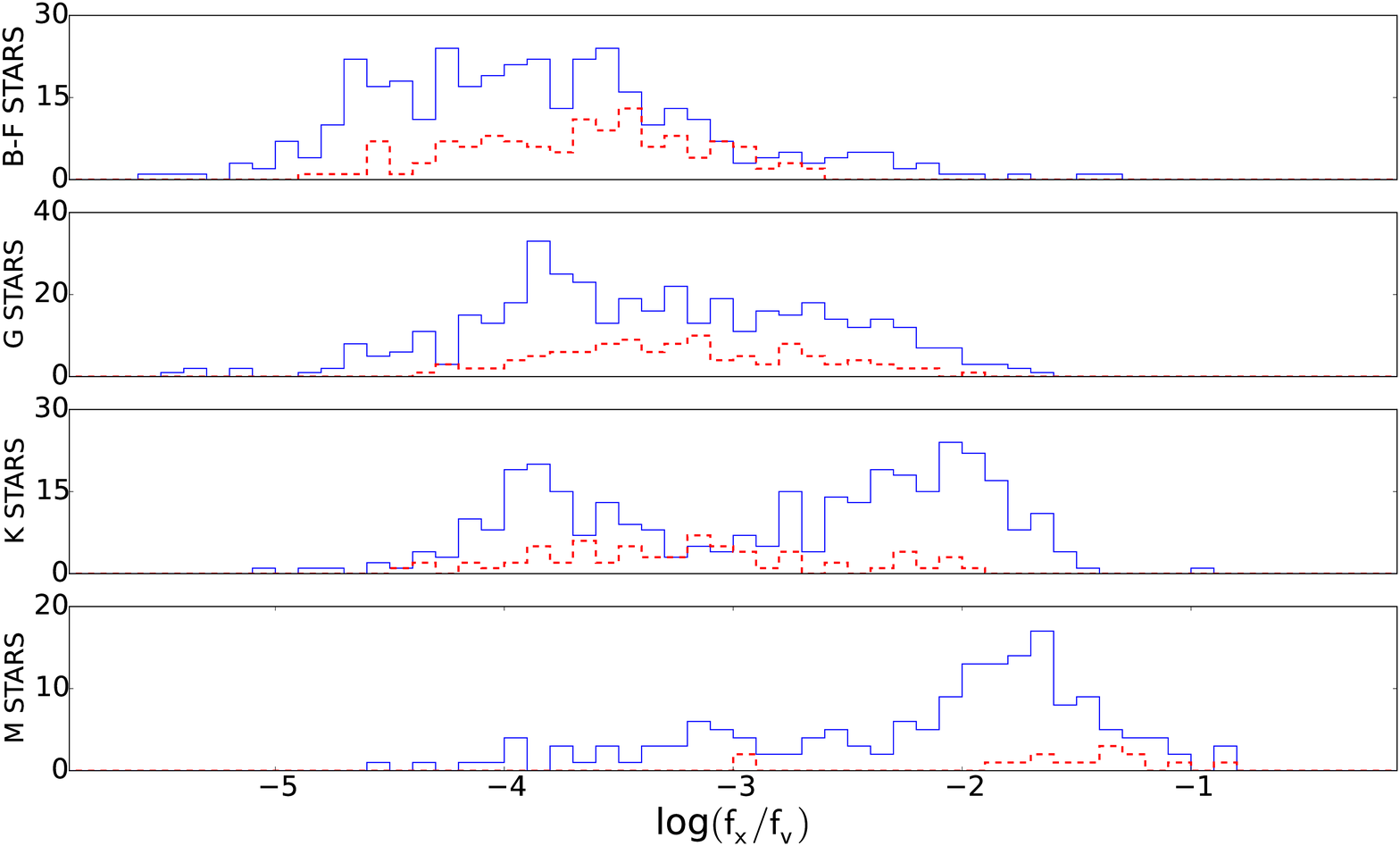}
\caption{Comparison of log($f_X/f_V$) distributions for each spectral type in our sample and in \citet{2009ApJS..181..444A}.
The red dashed histograms show the sample from \citet{2009ApJS..181..444A},
including 124, 110, 67 and 15 stars of the spectral types of F, G, K and M, respectively.
The blue histograms show the distributions of our sample, including two B stars, 36 A stars, 317 F stars, 405 G stars, 332 K stars, and 167 M stars.}
\label{spectral_compar.fig}
\end{figure*}

\citet{2009ApJS..181..444A} calculated log($f_X/f_V$) for 317 stars using the ROSAT All-Sky Survey \cite[RASS;][]{1999A&A...349..389V} and the SDSS catalog.
Figure \ref{spectral_compar.fig} shows a comparison of the log($f_X/f_V$) distribution between \citet{2009ApJS..181..444A} and our work.
There are 124, 110, 67, and 15 stars in the F, G, K, and M types from \citet{2009ApJS..181..444A}, respectively.
In our work, there are two B stars, 36 A stars, 317 F stars, 405 G stars, 332 K stars, and 167 M stars.

Generally, the log($f_X/f_V$) distributions of each spectral type in the two works are in good agreement.
However, the distribution of K type stars in our sample shows more obvious double-peak structure than that of \citet{2009ApJS..181..444A}.
There are much less active K stars in \citet{2009ApJS..181..444A}.
This could be due to the energy limit (0.2--2.4 keV) of the {\it ROSAT} mission,
which means that
\citet{2009ApJS..181..444A} may have lost some of the sources with harder spectra that have higher log($f_X/f_V$) values (Section \ref{hr.sec}).
On the other hand, there are more inactive M stars in our sample.
We propose this is due to a higher sensitivity of the {\it XMM-Newton} mission,
therefore more sources with lower log($f_X/f_V$) values can be detected.

\subsection{Correlation Between log($f_X/f_V$) and Stellar Parameters}
\label{pars.sec}

Using stellar parameter estimations from LAMOST, we study the correlations between log($f_X/f_V$) and stellar properties, including $T_{\rm eff}$, log($g$), [Fe/H], and H$\alpha$ emission.

\subsubsection{log($f_X/f_V$) vs. $T_{\rm eff}$ and log($g$)}
\label{fx2fot.sec}
%\par

\begin{figure}[htb]
%\figurenum{6}
\centering
\includegraphics[width=1.05\textwidth]{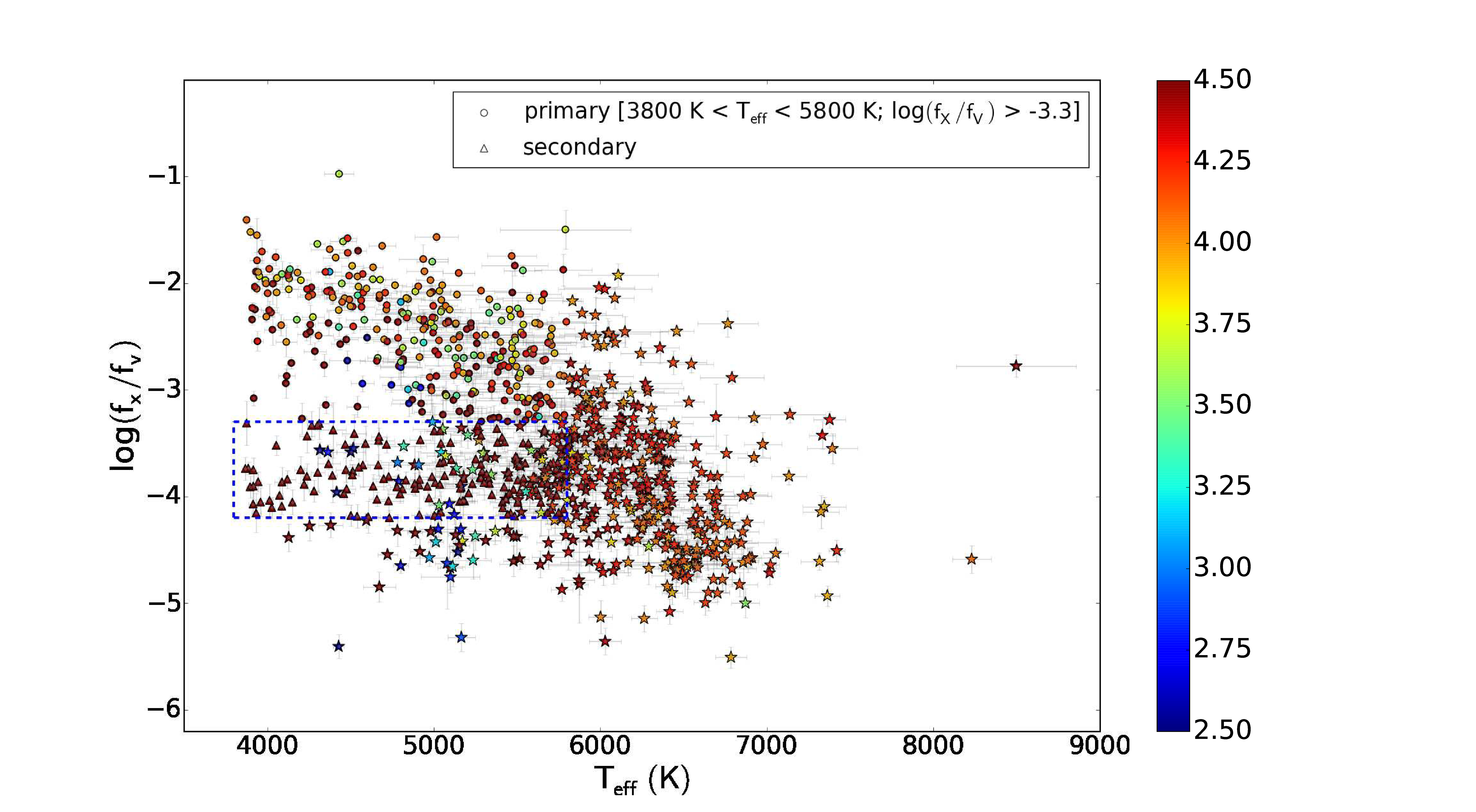}
\caption{
log($f_X/f_V$) as a function of temperature. The color shows different levels of $\log(g)$.
The triangles inside the dashed rectangle mark the sources of the secondary branch.
The circles represent the active part of the G and K stars in the primary branch, while the pentagrams represent the rest part of the primary branch.}
\label{T_ratio_G.fig}
\end{figure}

\begin{figure}[htb]
%\figurenum{6}
\centering
\includegraphics[width=1.05\textwidth]{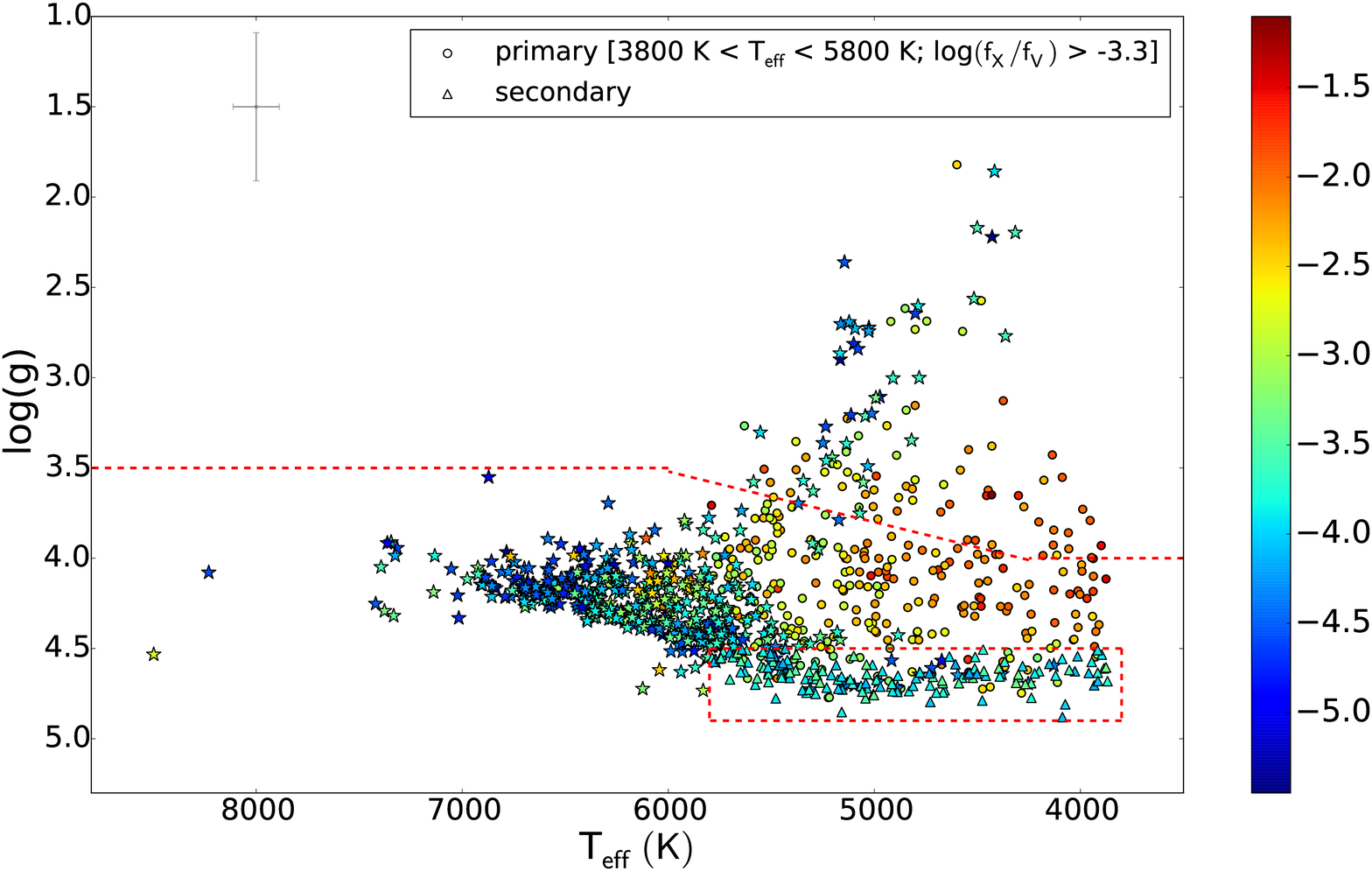}
\caption{
${\log(g)}$ as a function of temperature.
The color shows different levels of log($f_X/f_V$).
The red dashed line is the separation between giant and dwarf stars \citep{Ciardi2011}.
The meaning of the symbols is the same as in Figure \ref{T_ratio_G.fig}.
}
\label{T_G_ratio.fig}
\end{figure}

The bimodality of G- and K-type stars can also be seen in the log($f_X/f_V$)-$T_{\rm eff}$ diagram (Figure \ref{T_ratio_G.fig}).
We divided the sources into two branches.
For the primary branch, which contains the main part of the sources (outside the dashed rectangle),
log($f_X/f_V$) decreases with increasing temperature.
This result is compatible with previous works \citep[]{1991ApJS...76..813S,2009ApJS..181..444A}.
For the secondary branch (inside the blue dashed box),
stars generally have constant low log($f_X/f_V$) values for varying effective temperatures.
The secondary branch is mainly constructed of cool stars ($T_{\rm eff}$ $<$ 5800 K).
We roughly defined a region for the secondary branch in the stellar parameter space:
3800 K $<$ $T_{\rm eff}$ $\leq$ 5800 K;
$-$4.2 $<$ log($f_X/f_V$) $\leq$ $-$3.3;
4.5 $<$ log$(g)$ $\leq$ 4.9.
The constraint on log$(g)$ aims to exclude giant stars.
The secondary branch contains most of the inactive G- and K-type stars, and it can be regarded as the inactive part of the two types stars.
On the other hand, a group of stars (3800 K $< T_{\rm eff} < $ 5800 K; log($f_X/f_V$) $>$ $-$3.3) in the primary branch
can be considered as the active part of the G and K stars.

There are $\approx$ 108 giants showing X-ray emission, and some have high X-ray activity (Figure \ref{T_G_ratio.fig}).
This is consistent with previous studies that
late-type giants can have (high) stellar activities \citep{Simon1989, Auriere2015}.
However, some giants or sub-giants showing stellar activity may be in unrecognized binary systems \citep{Ozdarcan2018}, and our data are not sufficient to associate the X-ray emission to the giants themselves or their unresolved dwarf companions \citep{Schroder2007}.

\subsubsection{log($f_X/f_V$) vs. [Fe/H]}
%\label{xtime.sec}

\begin{figure*}[!htb]
%\figurenum{8}
\center
\includegraphics[width=1.05\textwidth]{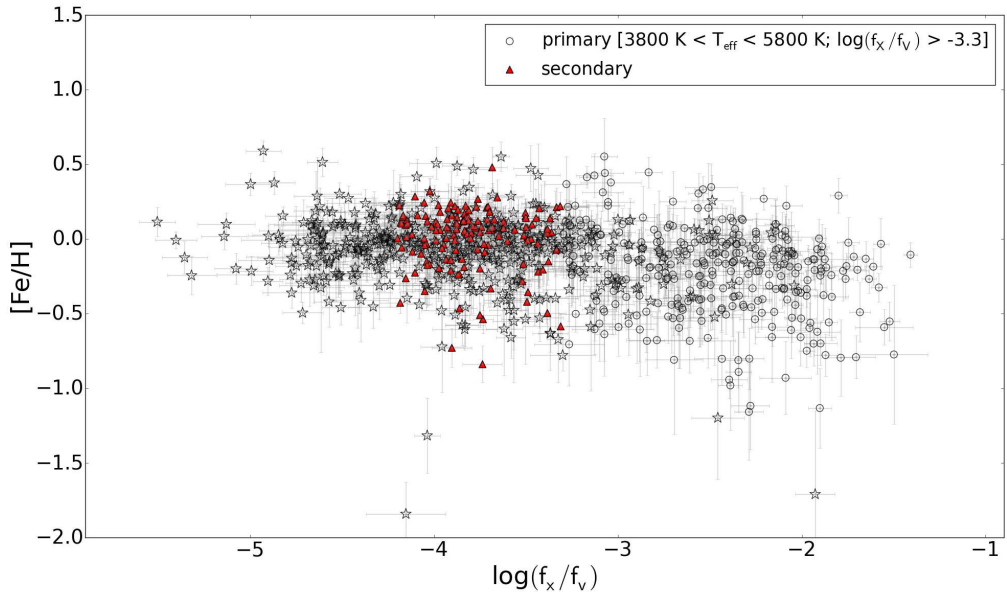}
\caption{Metallicity as a function of log($f_X/f_V$). 
The meaning of the symbols is the same as in Figure \ref{T_ratio_G.fig}.
%The red triangles are stars belonging to secondary branch,
%while the black circles are stars belonging to primary branch. 
}
\label{Fe_H_ratio.fig}
\end{figure*}

\begin{figure*}[!htb]
%\figurenum{10}
\center
\includegraphics[width=1.05\textwidth]{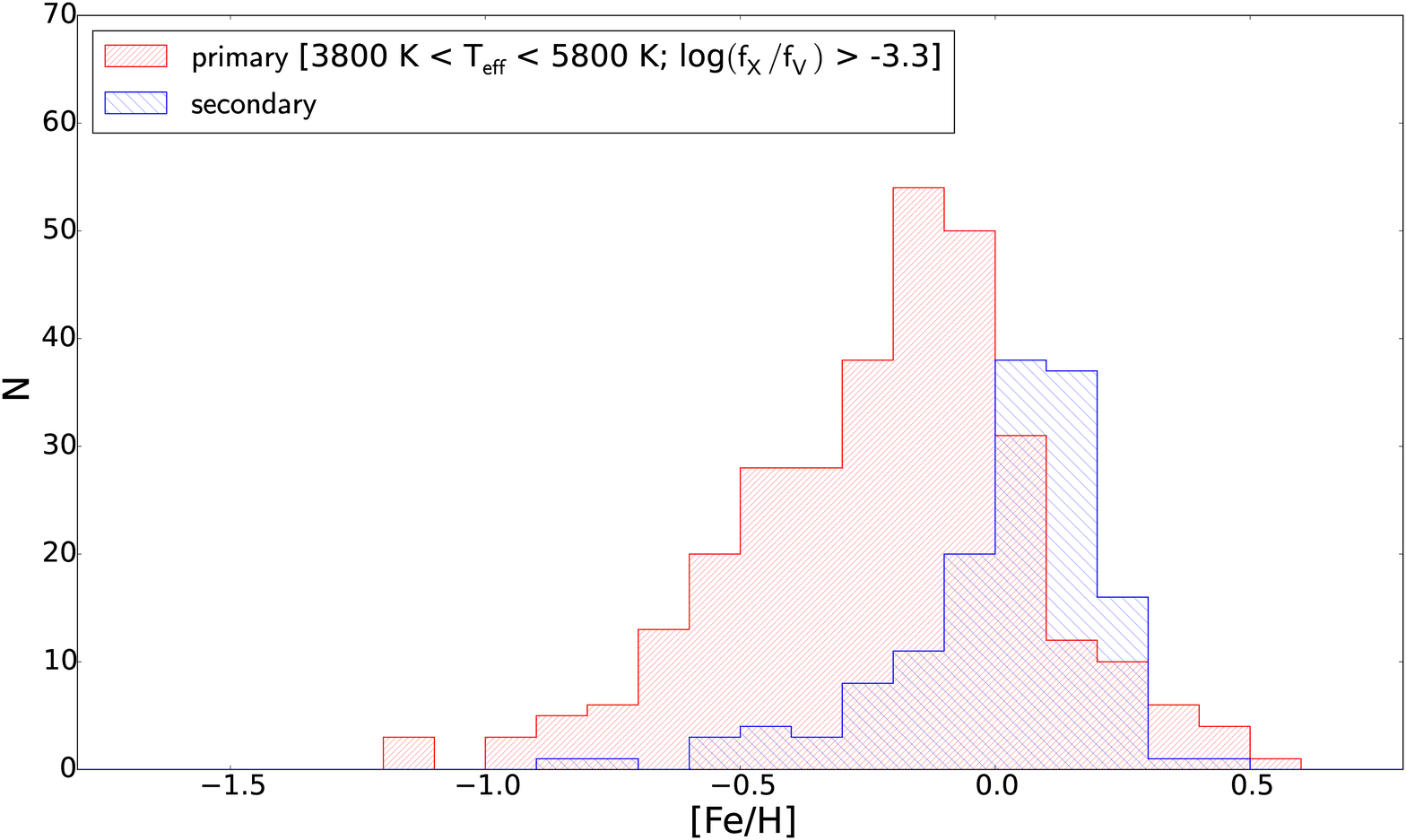}
\caption{Distribution of metallicities for stars in the primary and secondary branches.}
%The red and blue shadow region are the primary and secondary branch.}
\label{hist_feh.fig}
\end{figure*}

There is no clear evidence for a correlation between X-ray activity and metallicity (Figure \ref{Fe_H_ratio.fig}),
however, for active stars (log($f_X/f_V) > -$3),
a weak correlation between log($f_X/f_V$) and [Fe/H] is seen:
more active stars are more metal-poor.
%There is a weak correlation between metallicity and X-ray activity for the primary branch (Figure \ref{Fe_H_ratio.fig}):
%more active stars are more metal-poor.
This is consistent with \citet{1998MNRAS.298..332R},
who reported that the metallicities of very active stars are lower than those of normally active stars.
%Most of the stars with [Fe/H] $>$ 0.2 are located in the inactive region, for both the primary and secondary branch.
Those inactive stars in the secondary branch are generally more metal-rich than
the active stars (of the same spectral type) in the primary branch (Figure \ref{hist_feh.fig}).

\subsubsection{log($f_X/f_V$) vs. H$\alpha$ Emission}
%\label{xtime.sec}

\begin{figure*}[!htb]
%\figurenum{11}
\center
\includegraphics[width=1.05\textwidth]{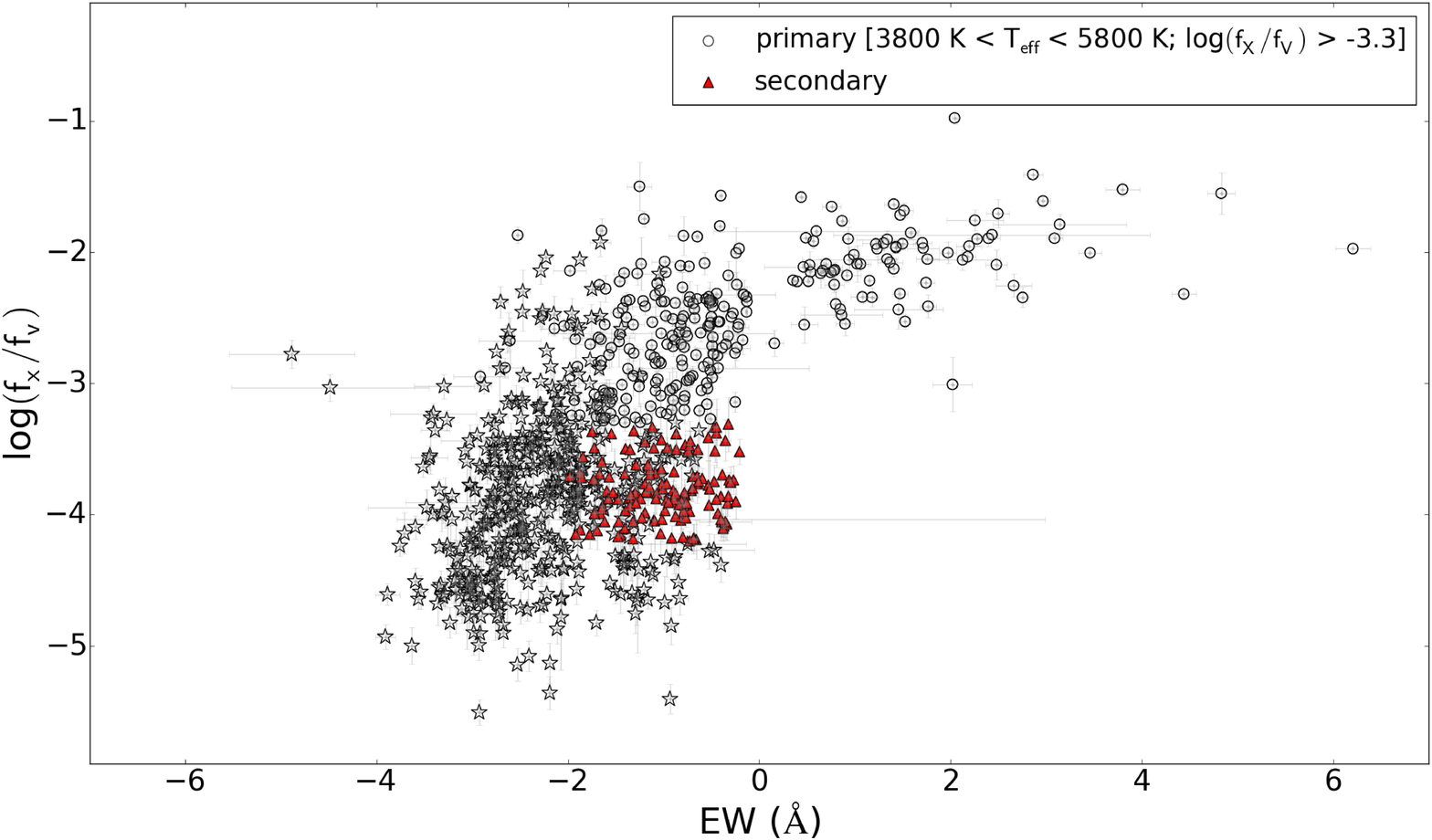}
\caption{%${\log(g)}$
log($f_X/f_V$) as a function of the EW of the H$\alpha$ lines.
The meaning of the symbols is the same as in Figure \ref{T_ratio_G.fig}.
%The red triangles and black circles are the sources belonging to the secondary and primary branch, respectively.
}
\label{ratio_EW.fig}
\end{figure*}

\begin{figure*}[!htb]
%\figurenum{7}
\center
\includegraphics[width=1\textwidth]{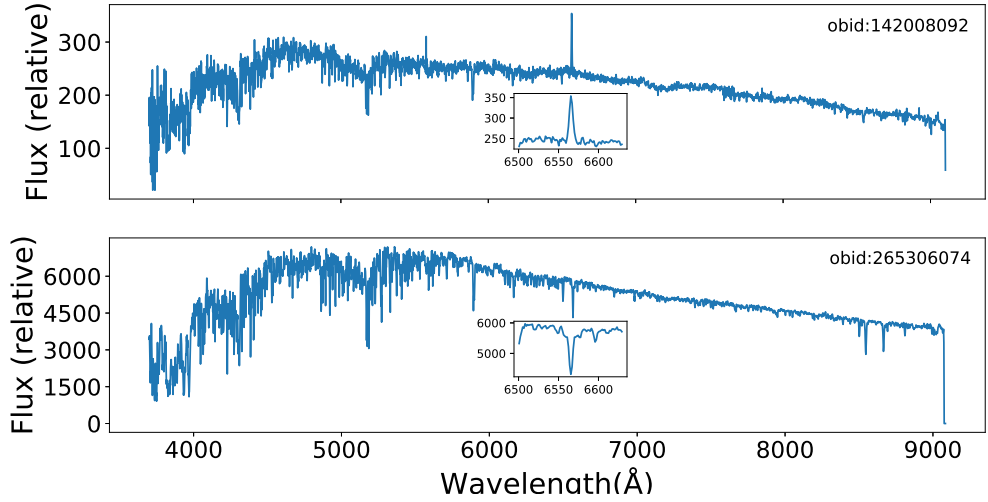}
\caption{Examples of LAMOST spectra for one active star (top panel) and one inactive star (bottom panel). The subplots show the H$\alpha$ lines.}
\label{act_spectra.fig}
\end{figure*}

Both the X-ray and H$\alpha$ emission are proxies of stellar magnetic activity \citep{Testa2015},
although they exist at different layers of the stellar atmosphere (i.e., corona and chromosphere).
LAMOST in combination with XMM-Newton provide us with a great opportunity to study the relation between the two activity indicators.
The EW of H$\alpha$ lines is listed in Table 3.
Stars with positive EW, which means H$\alpha$ emission line, have higher log($f_X/f_V$) values (Figure \ref{ratio_EW.fig}).
All stars in the secondary branch (i.e., X-ray inactive) and most of the stars in the primary branch,
which have low log($f_X/f_V$) values, do not have H$\alpha$ excess emissions.
As an example, the LAMOST spectra for two stars (one active and one inactive) are shown in Figure \ref{act_spectra.fig}.

\begin{figure*}[!htb]
%\figurenum{11}
\center
\includegraphics[width=1.05\textwidth]{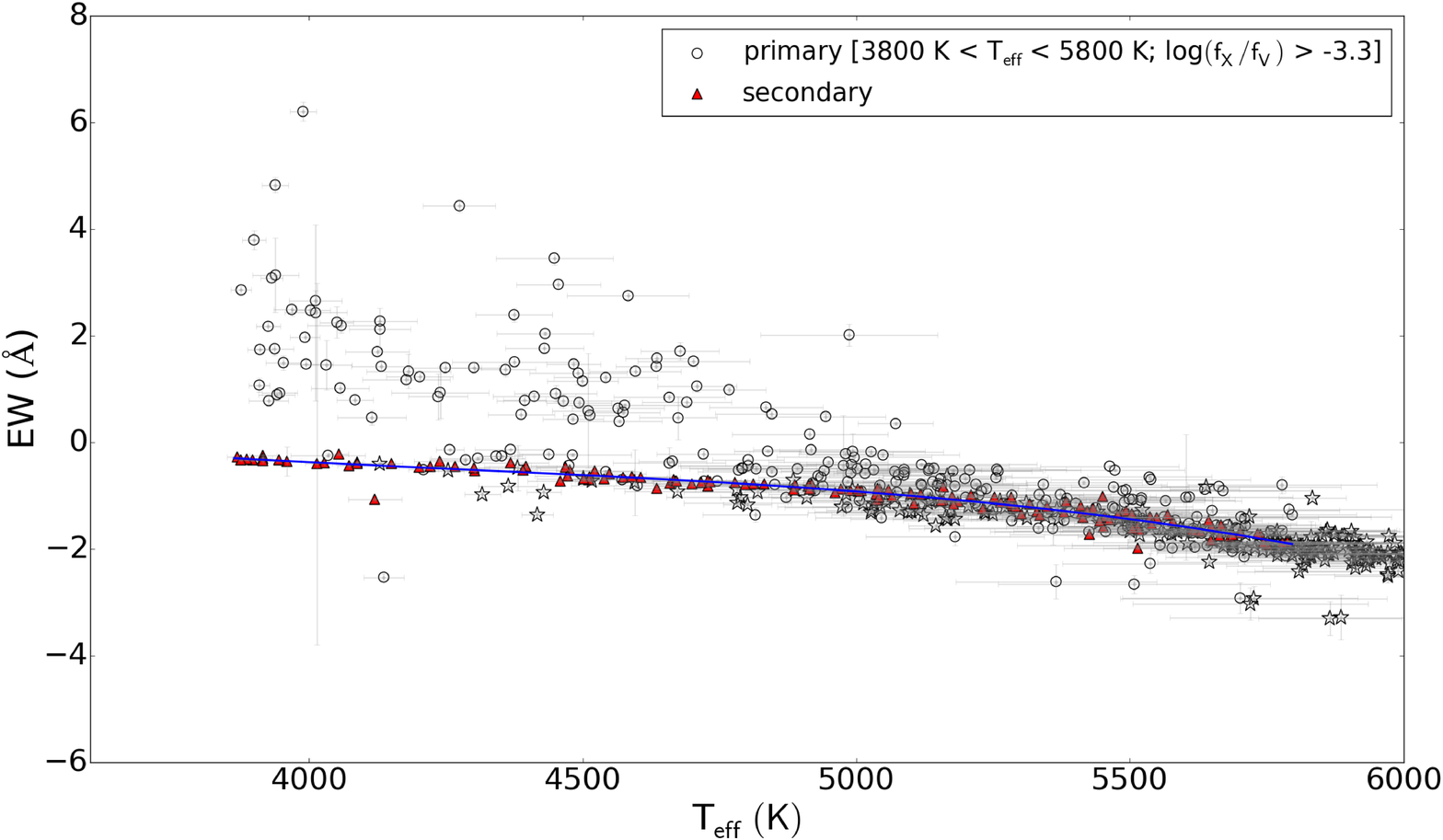}
\caption{EW$_{H\alpha}$ as a function of $T_{\rm eff}$. 
%The black circles are from the primary branch, and the red triangles are from the secondary branch. 
The meaning of the symbols is the same as in Figure \ref{T_ratio_G.fig}.
The solid line is the fitted ``basal line'' using the inactive stars in the secondary branch.}
\label{basal_EW.fig}
\end{figure*}

\begin{figure*}[!htb]
%\figurenum{13}
\center
\includegraphics[width=1.1\textwidth]{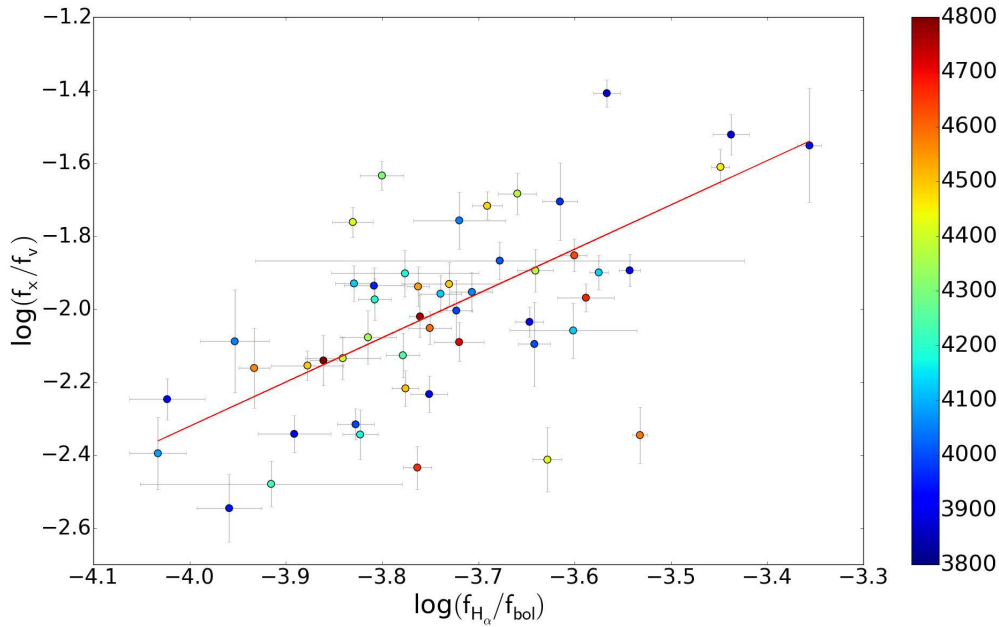}
\caption{log($f_X/f_V$) as a function of log($f_{H\alpha}/f_{\rm bol}$). The color shows different effective temperatures. The red line corresponds to the relationship
${\rm log}(f_X/f_V) = (1.21 \pm 0.23) \times {\rm log}(f_{H\alpha}/f_{\rm bol}) + (2.53 \pm 0.87)$.}
%${\rm log}(f_X/f_V) = (0.89 \pm 0.17) \times {\rm log}(f_{H\alpha}/f_{\rm bol}) + (1.44 \pm 0.68)$.}
%$log(f_X/f_V)$ = 0.89 $\times$ $log(f_{H\alpha}/f_{bol})$ + 1.44.}
\label{Ha_bol.fig}
\end{figure*}

However, EW is not a suitable indicator of stellar activity
since the continuum flux is very sensitive to the effective temperature \citep{Reid1995, 2017ApJ...849...36Y}.
To remove the effects of effective temperature and surface gravity,
we used the ratio of H$\alpha$ excess flux and bolometric flux to describe the activity of chromosphere.
First, we constructed a ``basal line'' of H$\alpha$ emissions using those inactive stars in the secondary branch (Figure \ref{basal_EW.fig}).
The excess EW (hereafter EW$'$) is calculated by subtracting the basal value at the same temperature, i.e.,
\begin{equation}
%{\rm EW}^{'}_{H\alpha} = {\rm EW}_{H\alpha} - {\rm EW}_{H\alpha,{\rm basal}}.
{\rm EW}^{'} = {\rm EW} - {\rm EW}_{\rm basal}.
\label{equ:exew}
\end{equation}
Then, we calculated the stellar surface fluxes of H$\alpha$ emission lines ($f_{\rm H\alpha}$) using the stellar atmosphere model CK04 \citep{ck04} based on the EW$'$.
The CK04 models list physical fluxes of the spectra in unit of ergs cm$^{-2}$ s$^{-1}$ A$^{-1}$.
For each star, the model with the most similar $T_{\rm eff}$, log$g$, and [Fe/H] was used.
Finally, we determined the flux ratio $f_{\rm H\alpha}/f_{\rm bol}$ using the bolometric flux $f_{\rm bol}=\sigma \,{T}^{4}$,
with the stellar temperature from LAMOST.
A power law dependence of $f_X/f_V$ on $f_{H\alpha}/f_{\rm bol}$ (Figure \ref{Ha_bol.fig})
is determined as,
%The best-fit line corresponds to the relationship:
\begin{equation}
{\rm log}(f_X/f_V) = (1.21 \pm 0.23) \times {\rm log}(f_{H\alpha}/f_{\rm bol}) + (2.53 \pm 0.87).
\end{equation}
%The tight correlation suggests that both chromospheric and coronal activity are independent of optical continuum (i.e., neither of them is a strong function of effective temperature).

Recently, \citet{Martinez-Arnaiz2011} reported a relation between X-ray and H$\alpha$ emission as $F_X \propto F_{\rm H\alpha}^{1.48\pm0.07}$,
using a sample of late-type dwarf active stars with spectral types from F to M.
For M dwarfs, \citet{Stelzer2013} derived $L_X/L_{\rm bol} \propto (f_{\rm H\alpha}/f_{\rm bol})^{1.90\pm0.31}$. In our study, we have obtained that $f_X/f_V \propto (f_{H\alpha}/f_{\rm bol})^{1.12 \pm 0.30}$, a slightly flatter relation than found in those other studies. The discrepancy may be due to our small sample size. In addition, we should note that the lack of simultaneous observations of those two intrinsically varying properties (coronal and chromospheric activities) may introduce another source of uncertainty in all these studies \citep{Martinez-Arnaiz2011}.

\section{The Bimodality of X-ray Activity}
\label{bimodality.sec}

The log($f_X/f_V$) distributions of G and K stars show clear bimodality (See Figure \ref{double_hist.fig} and \ref{T_ratio_G.fig}).
The gap between the peaks of the bimodal distribution is similar to that discovered by \citet{1980PASP...92..385V} using the Ca II H\&K lines. That gap was first explained as a sudden change of dynamo activity
to a less efficient mode at a critical rotation rate, but the scenario was rejected because the dependence of the chromospheric emission on rotation and spectral type is the same for stars above and below the gap \citep{1984ApJ...279..763N}.
Other studies with Ca II H\&K,  H$\alpha$, and X-ray emission
have also found the bimodality of stellar activity, with an active and inactive peak
\citep[e.g.,][]{Henry1996, 2004ApJS..152..261W, Jenkins2011,Martinez-Arnaiz2011,Pace2013}.
The bimodality is now explained as one young and one old subpopulation.
The old one is often thought to be inactive in chromospheric and X-ray emission, since
the magnetic activity decreases simultaneously as the rotation decelerates with age \citep[e.g.,][]{Mamajek2008,Katsova2011}.

\begin{figure*}[!htb]
%\figurenum{7}
\center
\includegraphics[width=1.05\textwidth]{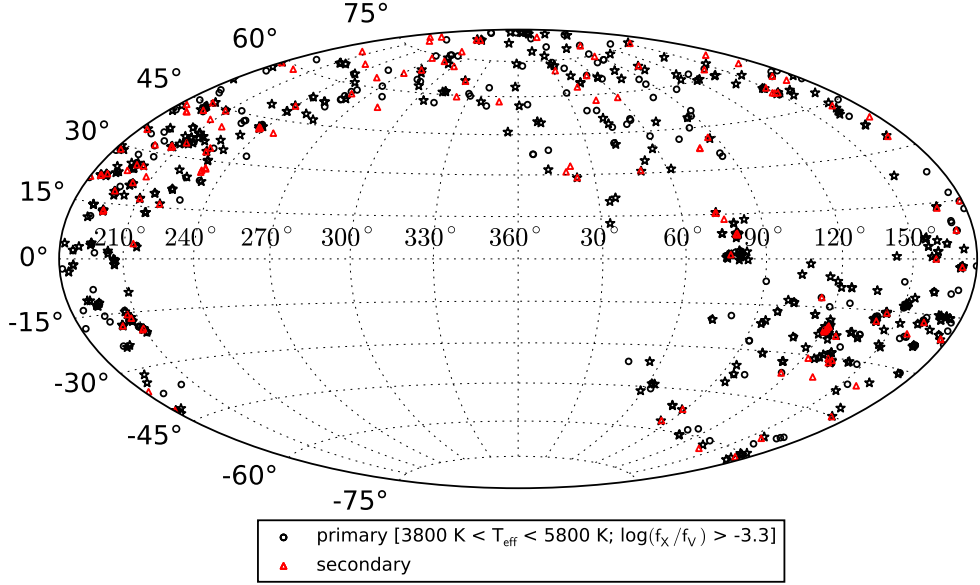}
\caption{The sky distribution of our stellar sources in Galactic coordinations.
The meaning of the symbols is the same as in Figure \ref{T_ratio_G.fig}.
%The red circles are stars belonging to secondary branch,
%while the blue squares are stars belonging to primary branch.
}
\label{sourceGal.fig}
\end{figure*}

The sky distribution (Figure \ref{sourceGal.fig}) shows no obvious distinction of the stars in the two branches.
That means those two branches are not belonging to different local structures (e.g., stellar streams).
Before we discuss the bimodality of X-ray activity,
we firstly checked the possibility that the bimodality is caused by selection effects, i.e.,
whether the detection limitations of {\it XMM-Newton} and LAMOST
can produce a double-peaked distribution from a single-peaked $f_X$ and a single-peaked $V$ distribution.

\subsection{Selection Effect}
%\label{xtime.sec}

\begin{figure*}[!htb]
%\figurenum{16}
\center
\includegraphics[width=1.1\textwidth]{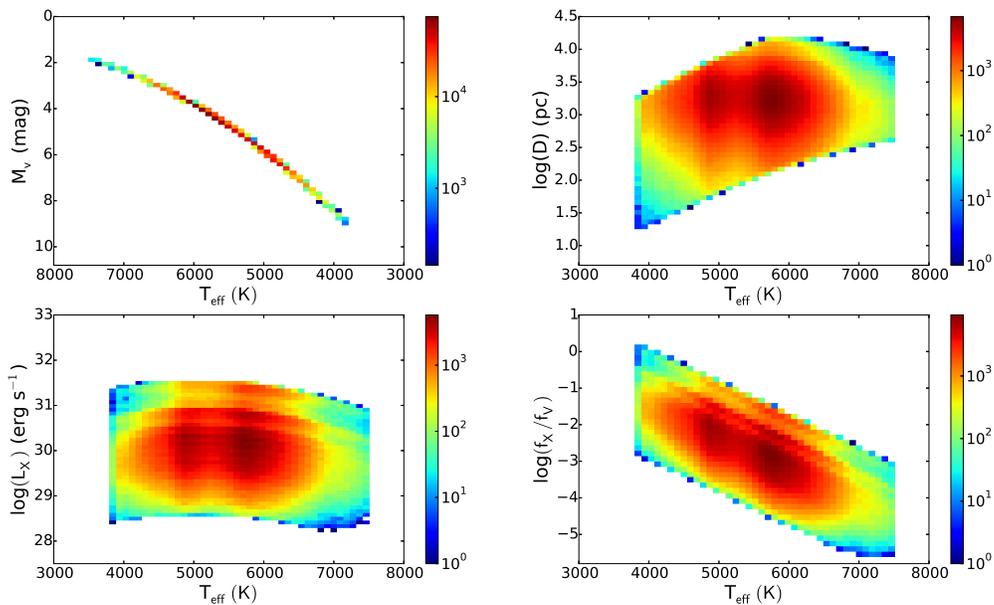}
\caption{Distributions of the parameters from the simulated sample.
The color indicates the numbers of simulated sources.
Top left panel: absolute magnitude as a function of temperature for the simulated sources.
Bottom left panel: X-ray luminosity as a function of temperature for the simulated sources.
Top right panel: distance as a function of the temperature for the simulated sources.
Bottom right panel: X-ray to optical flux ratio as a function of temperature for the simulated sources.}
\label{select_eff.fig}
\end{figure*}

We performed a Monte Carlo simulation to check whether the bimodality is caused by selection effects.
The detailed steps are as follows:\\
(i) We simulated a sample including 10 million sources with different effective temperatures,
using the proportion from the LAMOST stellar parameter catalogs.\\
(ii) We obtained the relation between log($f_X/f_V$) and temperature by fitting to the observed distribution of the primary branch as,
\begin{equation}
{\rm log}(f_X/f_V) = (-9.05 \pm 0.33) \times 10^{-4} T + (1.99 \pm 0.19).
\label{fx2fvt.eq}
\end{equation}
Using this Equation, we calculated log($f_X/f_V$) values for the 10 million simulated sources.\\
(iii) We obtained the relation between absolute magnitude and temperature by fitting the data in \citet{Wegner2007},
using a cubic polynomial equation as,
\begin{equation}
M_V = (-7.19 \pm 5.69) \times 10^{-12} T^{3} + (4.0 \pm 1.09) \times 10^{-7}T^{2} - (57.6 \pm 6.74) \times 10^{-4}T + (25.5 \pm 1.32).
\label{mv.eq}
\end{equation}
Using this Equation, we calculated the absolute magnitudes for the simulated sources (Figure \ref{select_eff.fig}, top left panel).\\
(iv) We derived the relation between X-ray luminosity and temperature using Equation \ref{fx2fvt.eq} and \ref{mv.eq}, following,
\begin{equation}
{\rm log}(L_X/L_V) = {\rm log}(f_X/f_V),
\label{lx2lv.eq}
\end{equation}
and
\begin{equation}
{\rm log}(L_V) = -0.4\times (M_V-M_{\odot,V}) - {\rm log}(L_{\odot,V}).
\label{lv.eq}
\end{equation}
The absolute magnitude and luminosity for the Sun is 4.85 mag \citep{1994ApJS...94..687W} and $4.64\times10^{32}$ erg s$^{-1}$
\footnote{http://astro.pas.rochester.edu/~aquillen/ast142/costanti.html}.
We then obtained the X-ray luminosities for the simulated stars (Figure \ref{select_eff.fig}, bottom left panel).\\
(v) Assuming the stars are located in the Galaxy,
we assigned random distances \citep[from 1 pc to 15 kpc;][]{2017A&A...602A..67A}) to each star (Figure \ref{select_eff.fig}, top right panel).\\
(vi) We calculated the apparent magnitude and X-ray flux using the simulated $M_V$, $L_X$, and distance $D$.
The detection limits were set for LAMOST ($10 < V < 20$) and {\it XMM-Newton} ($10^{-15}$ erg s$^{-1} < f_X < 10^{-11}$ erg s$^{-1}$) to select simulated sources that can be detected.\\
(vii) We re-calculated the log($f_X/f_V$) with Equation \ref{fx2fo.eq}, using the apparent magnitude and X-ray flux. The simulated distribution is shown in  Figure \ref{select_eff.fig} (bottom right panel).

\begin{figure*}[!htb]
%\figurenum{16}
\center
\includegraphics[width=0.49\textwidth]{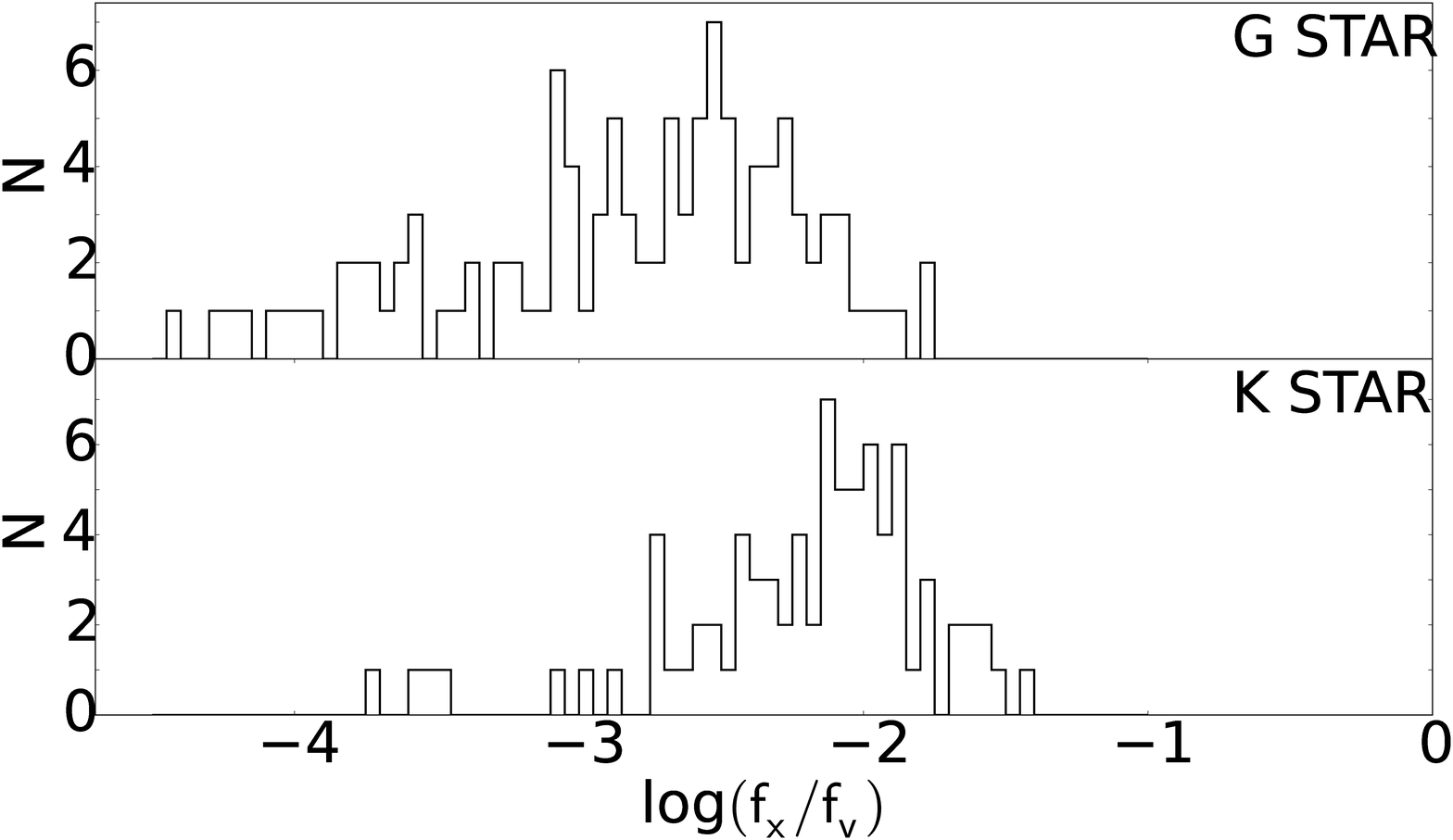}
\includegraphics[width=0.49\textwidth]{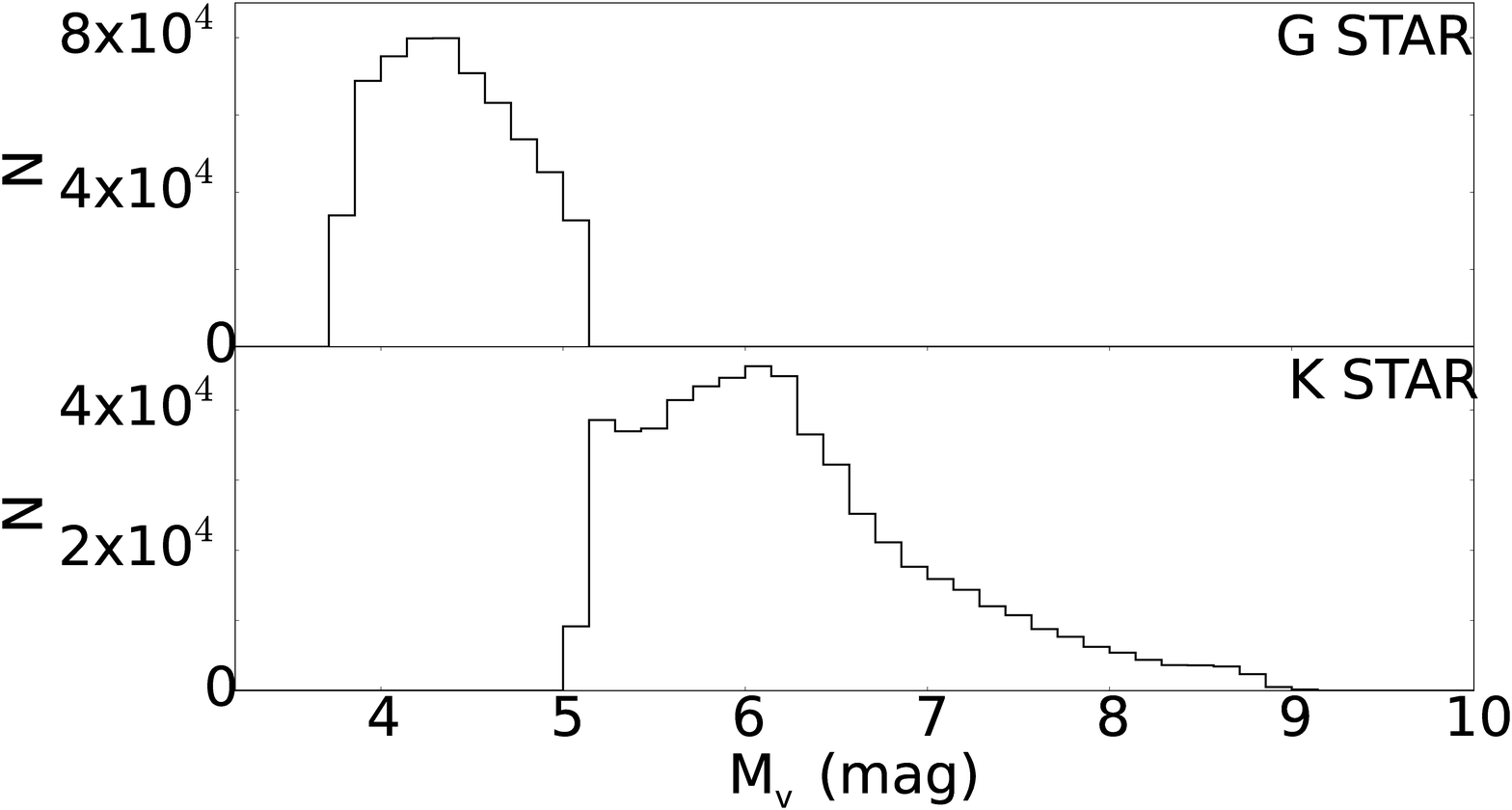}
\includegraphics[width=0.49\textwidth]{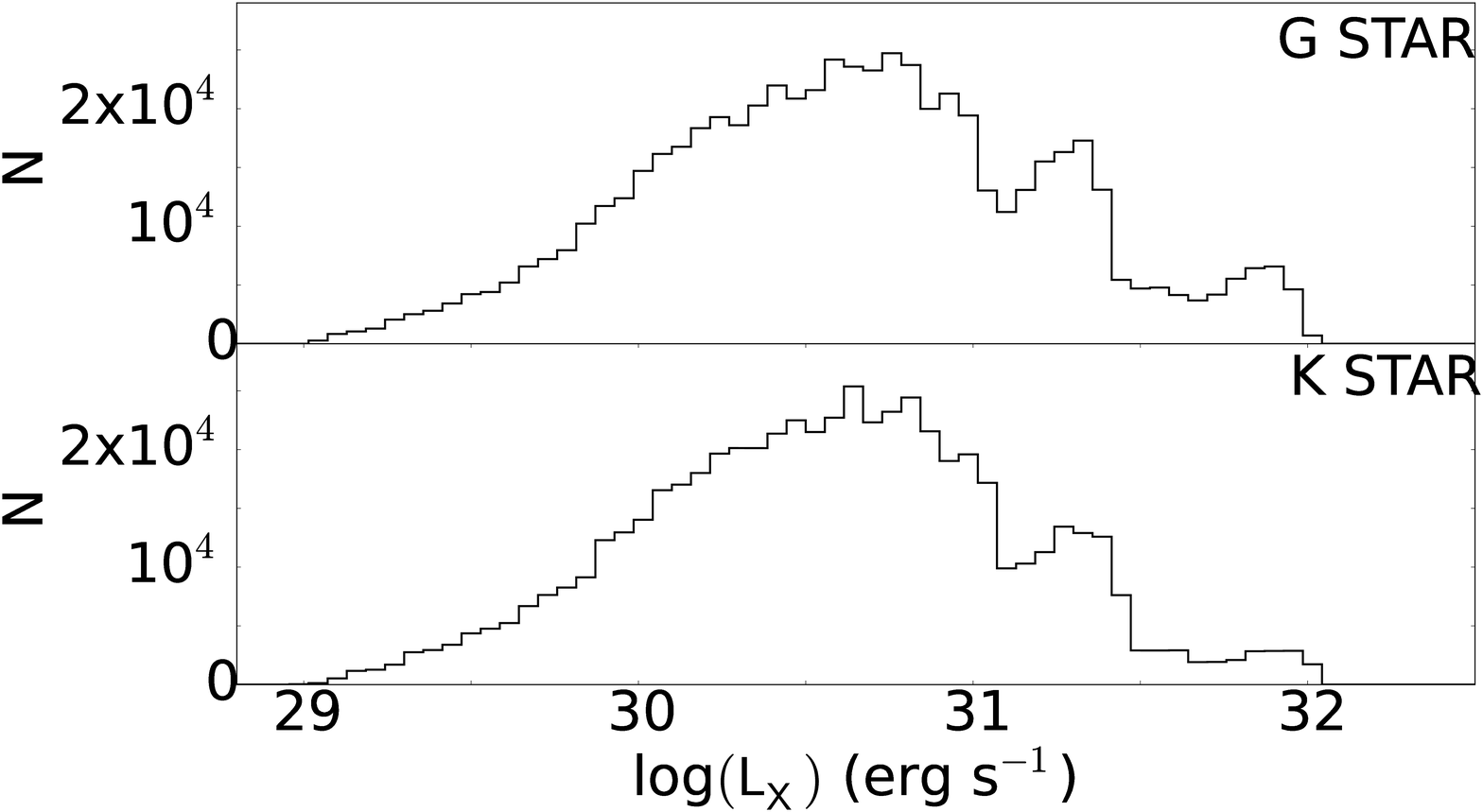}
\includegraphics[width=0.49\textwidth]{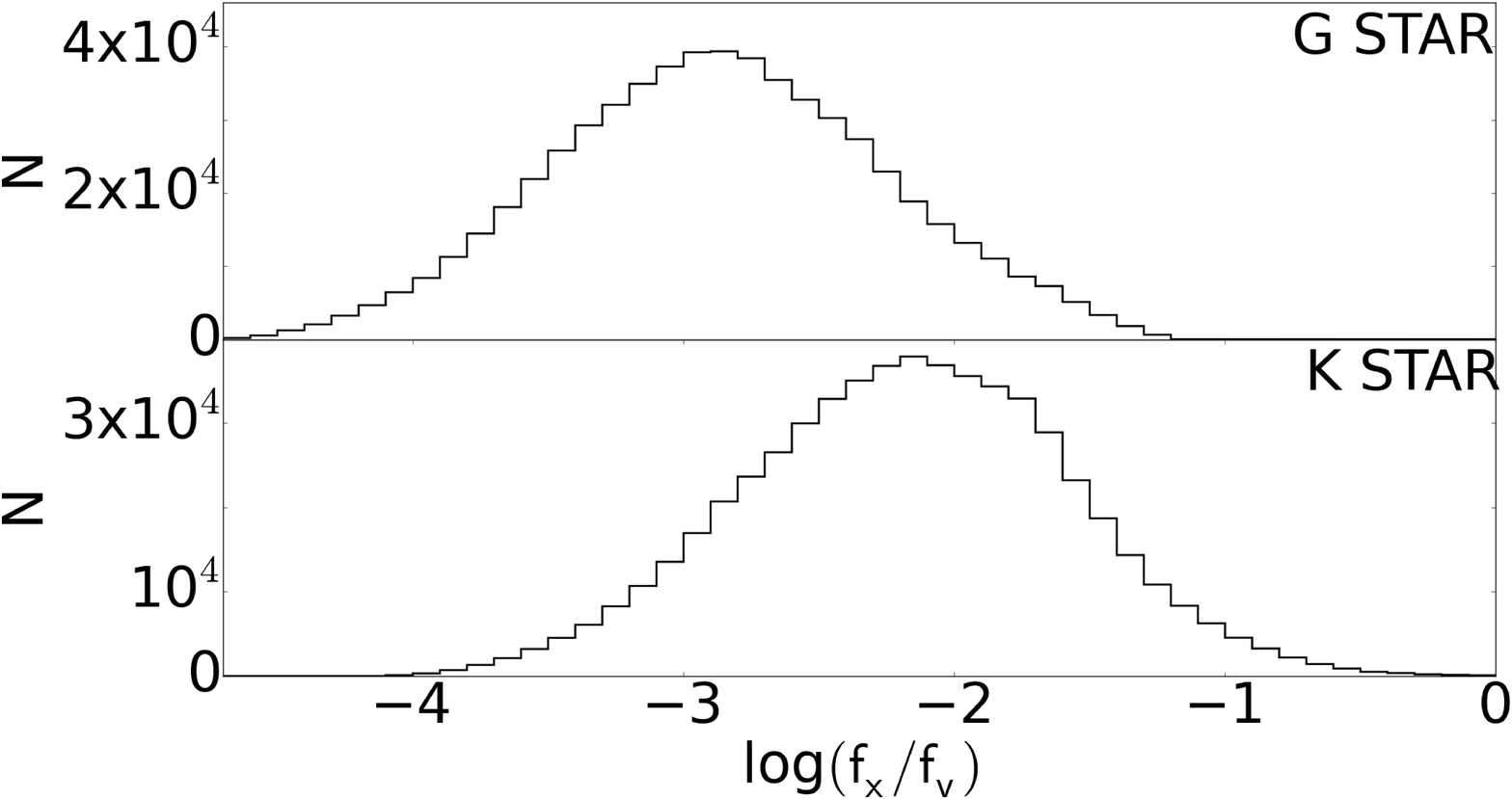}
\caption{Histograms of some parameters for G and K stars.
Top left panel: The observed log($f_X/f_V$) distributions of G and K stars in the primary branch.
Top right panel: The $M_V$ distributions for the simulated G and K stars.
Bottom left panel: The $L_X$ distributions for the simulated G and K stars.
Bottom right panel: The log($f_X/f_V$) distributions of the simulated G and K stars.}
\label{hist_sim.fig}
\end{figure*}

The simulated distributions of $M_V$, $L_X$, and log($f_X/f_V$) for G and K stars
are plotted in Figure \ref{hist_sim.fig}.
The simulated single-peaked distributions of log($f_X/f_V$) mean that
the observed bimodality of G and K stars (Figure \ref{double_hist.fig}) is not due to
 selection effect.

\subsection{$Hardness\ Ratio$}
\label{hr.sec}

The coronal temperature is known to be positively correlated with X-ray luminosity and stellar activity
\citep[e.g.,][]{Vaiana1983,Schrijver1984,Schmitt1997,2004A&ARv..12...71G, Jeffries2006,Telleschi2005,Telleschi2007}.
The cause of this relation between coronal temperature and luminosity can be that they are both functions of magnetic activity \citep{2004A&ARv..12...71G}. A more efficient dynamo inside active stars (for example because of faster rotation) produces stronger magnetic fields in the corona, and consequently a higher rate of field line reconnections and flares. This results both in a larger density of energetic electrons in the corona, and in higher temperatures.
%and the relation is explained as that
%increased magnetic activity leads to more frequent interactions of magnetic fields,
%and thus an increased coronal heating rate \citep{Gudel2004}.
Therefore, we suggest that the double-peaked distribution of log($f_X/f_V$)
represents a double-peaked distribution of heating rates, and therefore coronal temperatures.

\begin{figure*}[!htb]
%\figurenum{14}
\center
\includegraphics[width=1.1\textwidth]{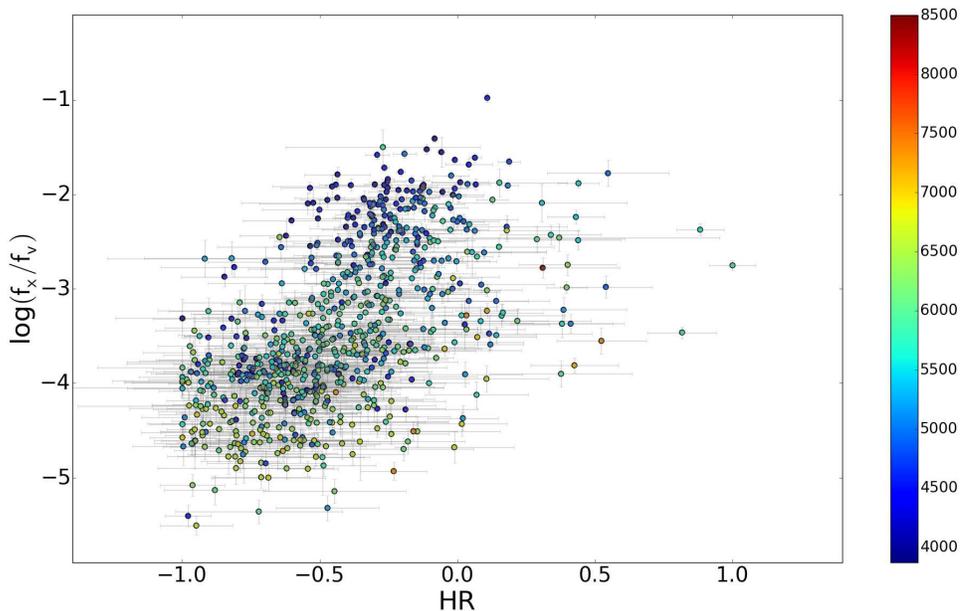}
\caption{%${\log(g)}$
log($f_X/f_V$) as a function of HR. The color shows different effective temperatures.}
\label{HR_2.fig}
\end{figure*}

\begin{figure*}[!htb]
%\figurenum{15}
\center
\includegraphics[width=1.05\textwidth]{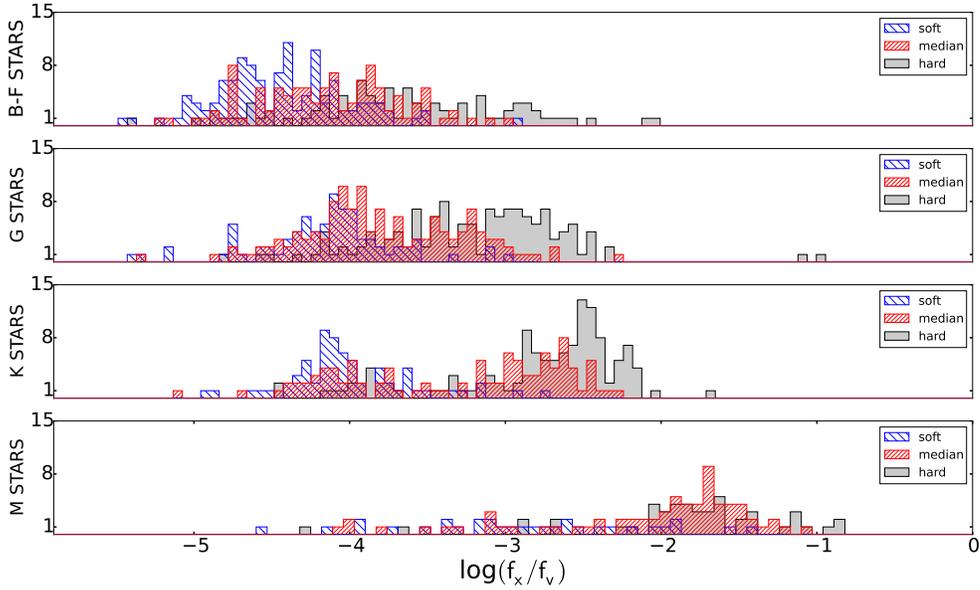}
\caption{Distribution of log($f_X/f_V$) for three types with different $HR$:
hard (${ 0.3 < HR < 1}$), median (${ -0.6 < HR < 0.3}$), and soft (${ -1 < HR < -0.6}$).
It is clear that hard sources have the highest log($f_X/f_V$) values, while soft sources have the lowest values.
The mean log($f_X/f_V$) values of the soft, median, and hard sources in the G type are $-$3.93, $-$3.48, and $-$2.81, and their spread around the mean log($f_X/f_V$) are 0.53, 0.61, and 0.69, respectively.
The mean log($f_X/f_V$) values of the three kinds of sources in the K type are $-$3.75, $-$2.91, and $-$2.33, and their spread around the mean log($f_X/f_V$) are 0.48, 0.8, and 0.67, respectively.}
%different hardness ratio in different spectral types. We defined the range of hardness ratio: soft is from -1 to -0.6; median is from -0.6 to 0.3; hard is from 0.3 to 1.}
\label{hist_HR.fig}
\end{figure*}

In our work, we take the X-ray $HR$ as a proxy for the coronal temperature,
because hotter corona will emit photons with higher energies, which produces harder X-ray spectra.
There is a positive correlation between log($f_X/f_V$) and $HR$ (Figure \ref{HR_2.fig}).
That means stronger X-ray emitters (higher log($f_X/f_V$)) have higher coronal temperatures.
To have a better look, we classified the objects into three types:
hard, median, soft sources, using the criteria as:
hard (${ 0.3 < HR < 1}$), median (${ -0.6 < HR < 0.3}$), and soft (${ -1 < HR < -0.6}$).
We find (Figure \ref{hist_HR.fig}) that
(1) the higher activity peak
of the log($f_X/f_V$) distribution in G-type stars is dominated by hard sources,
and the lower activity peak by median and soft sources;
(2) the higher activity peak
of the log($f_X/f_V$) distribution in K-type stars is dominated by hard and median sources,
and the lower activity peak by soft sources.

Besides G and K stars, the bimodality was also detected in late F and early M stars (Section \ref{fx2fv.sec}), although
the statistics are too poor due to the sample limit.
One question is that why the bimodality is not detected for those hotter or cooler stars.
For hotter stars (early F and earlier), the coronal heating efficiency may be quite low for most of them. For cooler stars (late M), firstly, they are generally optically faint, thus our sample may be not complete; secondly, stars later than M4 type may have different dynamo mechanism due to their fully convective feature \citep{Durney1993}; thirdly, the evolution of M stars is very slow, which means most of them are still in the regime with high coronal heating rate. In fact, most of the M stars have high HR values around 0 (Figure \ref{HR_2.fig}).

\begin{figure*}[!htb]
%\figurenum{19}
\center
\includegraphics[width=0.9\textwidth]{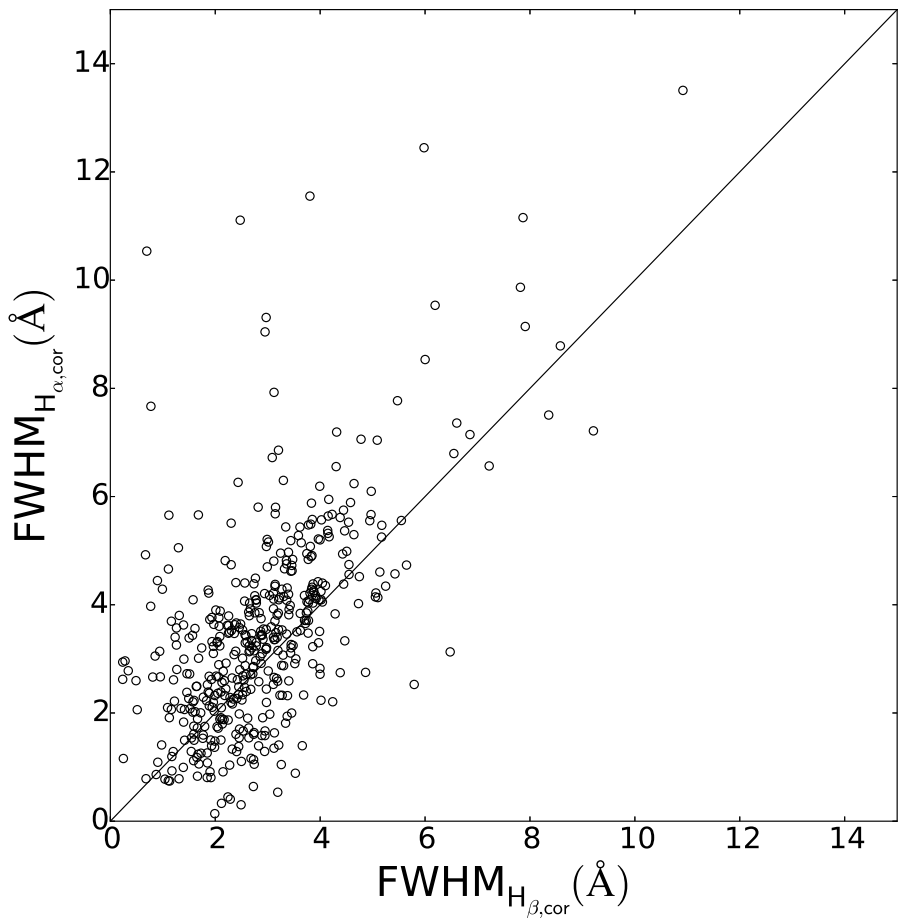}
\caption{Comparison of corrected FWHMs calculated from H$\alpha$ and H$\beta$ lines.}
\label{differ_total_vsini.fig}
\end{figure*}

\begin{figure*}[!htb]
%\figurenum{19}
\center
\includegraphics[width=1.05\textwidth]{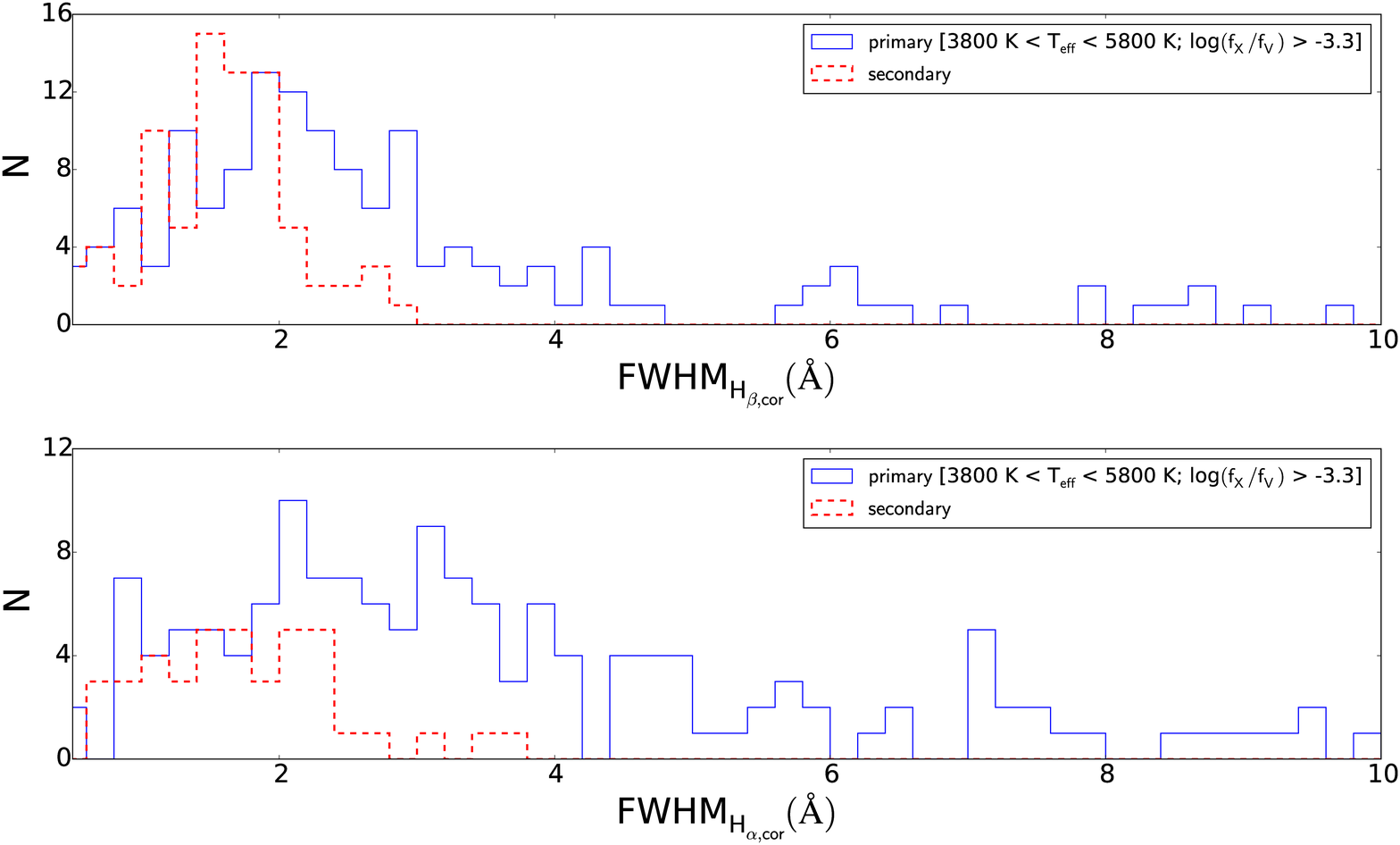}
\caption{Top panel: distributions of corrected FWHMs calculated from H$\beta$ lines for the primary and secondary branch.
Bottom panel: distributions of corrected FWHMs calculated from H$\alpha$ lines for the primary and secondary branch.}
%the number of sources in the meddle and bottom panel are normalized.}
\label{total_vsini.fig}
\end{figure*}

\subsection{Rotational Velocities}
\label{RV.sec}

%In general, X-ray emission from late-type stars is attributed to the presence of a magnetic corona.
%It is believed that the generation of magnetic energy by large-scale dynamo action
%is driven by rotation and convection \citep[e.g.,][]{Charbonneau2010, Reiners2014}.
The magnetic dynamo suggests a lower rotation velocity for inactive stars than active ones.
Therefore, we further checked the difference of rotational velocities of the primary branch (3800 K $< T_{\rm eff} < $ 5800 K; log($f_X/f_V$) $>$ $-$3.3) and secondary branch.
They can be regarded as the active and inactive parts (of G and K stars), respectively.
In our work, we take the FWHMs of Balmer lines as proxies of rotational velocities \citep{Strassmeier1990,Fekel1997}.
%to explain the difference between the primary and secondary branch,
%it is necessary to check the rotational velocities which can be expressed by the FWHM of certain lines \citep{Strassmeier1990,Fekel1997}.
%\citet{Strassmeier1990} reported the relation between the projected rotational velocity and the FWHM of several lines:

The instrumentally corrected FWHM is calculated as
\begin{equation}
{\rm FWHM}_{\rm cor} = ({\rm FWHM}^{2}_{\rm obs} - {\rm FWHM}_{\rm inst}^{2})^{1/2},
\end{equation}
where ${\rm FWHM_{obs}}$ is the observed FWHM, ${\rm FWHM_{inst}}$ the FWHM of the lamp lines, and ${\rm FWHM}_{\rm cor}$ the corrected FWHM
\citep{Strassmeier1990}.
Here we calculated the FWHMs of H$\alpha$ and H$\beta$ lines, for 540 and 596 stars, respectively (Table 3).
Collisional broadening and broadening as a result of macro-turbulence were not corrected for.
The corrected FWHMs measured from H$\alpha$ and H$\beta$ lines are in good agreement (Figure \ref{differ_total_vsini.fig}).
%It is clear that the distribution of FWHM measured from two lines are in good agreement.
The mean deviation between the FWHMs from the two lines is $\approx$0.43 {\AA} , with a standard deviation as $\approx$1.09 {\AA}.

The FWHMs of the secondary branch stars are generally smaller than those of the part of primary branch stars (Figure \ref{total_vsini.fig}),
indicating relatively lower rotational velocities of the secondary branch sources.
This further explains the bimodality of the X-ray activity of G and K stars.
Low rotational velocity weakens the coronal activity \citep{1981ApJ...248..279P, Pizzolato2003, 2011ApJ...743...48W} and, therefore the X-ray to optical flux ratio.

\section{CONCLUSION}
\label{summary.sec}

The {\it XMM-Newton} and LAMOST data allow us to identify X-ray emitters and probe stellar X-ray activity over a wide range of stellar parameters.
By cross matching the 3XMM-DR5 catalog and the LAMOST DR3 catalog,
we provide a sample including 1259 X-ray emitting stars,
of which 1090 have accurate stellar parameter estimations.
Our sample size is much larger than those in previous works.
%
%Based on the classifications from LAMOST, we provided a sample of X-ray sources including 134 galaxies, 60 QSOs, and 1259 stars.
%
We studied the X-ray emission level, using X-ray to optical flux ratio, for different stellar types, including two B stars, 36 A stars, 317 F stars, 405 G stars, 332 K stars, and 167 M stars.
%The X-ray to optical flux ratio, which represents X-ray activity, of late-type stars are clearly higher than those of early-type stars.
Late type stars in general have higher log($f_X/f_V$) values than early type ones, indicating their higher X-ray activity.

We find a bimodal distribution of log($f_X/f_V$) for G- and K-type stars.
We performed a Monte Carlo simulation which
proves that the double-peaked distribution are not caused by selection effect.
We explain this bimodality as evidence of two subpopulations with different coronal heating rates, and therefore different coronal temperatures.
Stars with a hotter corona --- observationally with a higher $HR$ --- have a higher X-ray log($f_X/f_V$) value.
Furthermore, we calculated the FWHMs of H$\alpha$ and H$\beta$ lines,
and found that those inactive stars have generally smaller FWHMs, and therefore lower rotational velocities, than active stars.
In fact, the log($f_X/f_V$) distributions of the late F and early M types also display weak bimodality, but the statistics are very poor due to the sample limit. Future studies with more F and M stars may shed more light on the distribution.
In general, the rotation velocity and stellar activity observationally decay with stellar age
\citep[e.g.,][]{Wilson1963, Skumanich1972, Simon1985, Cardini2007, Mamajek2008},
therefore, those inactive stars with much higher log($g$) values are possibly old stars.
We speculate that the old age, stellar activity cycles like those displayed by the sun \citep{Judge2003}, and long-term variation
\citep[such as Maunder minimum:][]{Baliunas1990, Henry1996} may all contribute to the inactive part.
We also examined the correlation between log($f_X/f_V$) and H$\alpha$ emission line luminosity,
and find a positive tight correlation between the two quantities.

%\begin{threeparttable}
%\begin{sidewaystable}[h]
\begin{table}
\renewcommand{\arraystretch}{1.5}
\tabcolsep=3pt
\scriptsize
\begin{center}
\centering \caption[]{The X-ray information for the sample sources.}
\label{xray.tab}
\vspace{0.2cm}
\begin{tabular}{cccc}
\hline
%obsid&RA&DEC&subclass&$f_X$&e$f_X$&${\log(f_X/f_V)}$&e${\log(f_X/f_V)}$&$HR$&e$HR$ \\
obsid&$f_X$&${\log(f_X/f_V)}$&$HR$ \\
     & (erg s$^{-1}$ cm$^{-2}$) & &  \\
(1)  &  (2)&   (3)  &(4) \\
\hline
1603040&9.67e-15$\pm$1.71e-15&-2.51$\pm$0.14&0.57$\pm$1.11 \\
7003074&5.06e-14$\pm$4.81e-15&-1.93$\pm$0.06&0.23$\pm$0.44 \\
7003243&5.11e-15$\pm$6.01e-16&-2.78$\pm$0.07&-1.00$\pm$2.41 \\
7711046&3.72e-14$\pm$3.63e-15&-2.14$\pm$0.06&-1.00$\pm$0.35 \\
20001189&5.74e-14$\pm$1.86e-15&-1.49$\pm$0.04&-0.36$\pm$0.06 \\
28306100&5.44e-15$\pm$8.75e-16&-3.89$\pm$0.08&-0.89$\pm$1.00 \\
51007061&1.68e-14$\pm$1.37e-15&-3.80$\pm$0.05&-1.00$\pm$0.90 \\
53111238&2.26e-14$\pm$2.53e-15&-4.12$\pm$0.06&-0.93$\pm$1.66 \\
74309103&1.69e-14$\pm$1.78e-15&-4.61$\pm$0.09&-1.00$\pm$0.53 \\
74311158&3.54e-14$\pm$2.38e-15&-4.19$\pm$0.08&-0.91$\pm$0.43 \\
74403036&1.74e-14$\pm$1.10e-15&-3.64$\pm$0.05&-0.55$\pm$0.29 \\
75805212&3.10e-14$\pm$1.48e-15&-2.42$\pm$0.05&-0.69$\pm$0.13 \\
76406132&2.17e-14$\pm$2.01e-15&-4.43$\pm$0.05&-0.82$\pm$0.40 \\
76506084&1.29e-14$\pm$1.32e-15&-3.59$\pm$0.06&-0.95$\pm$0.62 \\
84812137&2.92e-14$\pm$2.97e-15&-3.53$\pm$0.06&-0.74$\pm$0.94 \\

\hline
\end{tabular}
\end{center}
Note: Column 1: spectral ID in LAMOST catalog.
 Column 2: unabsorbed flux in band 0.3--3.5 keV. Column 3: X-ray to optical flux ratio.
 Column 4: EPIC $HR$ using the bands of 0.5--1 keV and 1--2 keV derived from 3XMM-DR5 catalog.
%\end{threeparttable}
\end{table}
%%\end{table} \footnote{lala}
%\end{sidewaystable}

%\begin{threeparttable}
\begin{table}
%%\begin{table}
\renewcommand{\arraystretch}{1.5}
\tabcolsep=3pt
\scriptsize
\begin{center}
\caption[]{Stellar parameters for the sample sources.}
\label{measure.tab}
\vspace{0.2cm}
\begin{tabular}{ccccccccccc}
\hline
obsid&RA&DEC&subclass&$V$&$A_V$&Nh&$T_{\rm eff}$&log($g$)&[Fe/H] \\
     & (deg)  &  (deg) & &(mag)&(mag)&(cm$^{-2}$)&(K)&   &     \\
(1)  &  (2)&   (3)  &(4)&(5)&(6)&(7)&(8)&(9)&(10) \\
\hline
1603040&9.873834&40.083248&G9&15.77$\pm$0.29&0.38&1.09e+21&5020.89$\pm$211.08&4.27$\pm$0.63&0.09$\pm$0.28 \\
1615104&11.123160&41.359930&K1&14.17$\pm$0.03&1.12&3.22e+21&4808.98$\pm$106.97&3.05$\pm$0.58&-0.58$\pm$0.16 \\
1615179&10.716530&41.518950&F6&15.81$\pm$0.01&1.31&3.77e+21&5280.53$\pm$322.29&3.04$\pm$1.17&-1.42$\pm$0.52 \\
1615187&10.939620&41.616010&K3&15.34$\pm$0.08&0.64&1.82e+21&4872.15$\pm$256.63&3.42$\pm$0.61&-0.24$\pm$0.33 \\
7003074&9.322592&40.763245&K5&15.25$\pm$0.05&0.23&6.54e+20&4491.66$\pm$88.32&4.22$\pm$0.26&-0.14$\pm$0.13\\
7003243&9.630176&40.286594&G8&15.70$\pm$0.07&0.30&8.58e+20&5141.44$\pm$194.22&4.08$\pm$0.54&-0.07$\pm$0.24\\
7711046&16.773656&32.182487&K5&15.02$\pm$0.05&0.19&5.32e+20&4464.46$\pm$121.34&4.01$\pm$0.30&0.10$\pm$0.16 \\
20001189&185.349817&28.070038&K3&16.15$\pm$0.01&0.17&4.74e+20&4809.03$\pm$154.27&4.64$\pm$0.27&-0.04$\pm$0.23 \\
28114180&98.454761&17.791983&G5&11.83$\pm$0.03&0.68&1.96e+21&5094.17$\pm$114.29&2.73$\pm$0.59&0.01$\pm$0.16 \\
28306100&139.828255&30.428329&G3&12.63$\pm$0.05&0.07&2.09e+20&5728.46$\pm$166.85&4.41$\pm$0.52&-0.10$\pm$0.18 \\
42815236&98.925573&5.526181&K0&11.61$\pm$0.14&0.89&2.55e+21&4745.80$\pm$54.52&2.69$\pm$0.52&-0.39$\pm$0.09 \\
43810009&98.397596&17.657734&K4&14.40$\pm$0.04&1.88&5.38e+21&4599.58$\pm$120.03&1.82$\pm$0.57&-0.66$\pm$0.20 \\
51007061&239.800512&27.268972&G3&11.75$\pm$0.01&0.20&5.81e+20&5821.85$\pm$122.08&4.06$\pm$0.56&0.06$\pm$0.13 \\
53111238&235.942430&54.151718&F6&10.58$\pm$0.04&0.16&4.57e+20&6404.79$\pm$121.72&4.19$\pm$0.42&0.03$\pm$0.12\\
74309103&35.130959&-6.186942&F0&9.61$\pm$0.17&0.09&2.47e+20&6750.24$\pm$182.92&4.13$\pm$0.38&-0.04$\pm$0.18 \\

\hline
%\tablecomments{Note that does not work with the
%vertical line alignment token. If you want vertical lines in the headers you
%can not use this command at this time.}
%\tablecomments{Note that {\tt \string \colnumbers} does not work with the
%vertical line alignment token. If you want vertical lines in the headers you
%can not use this command at this time.}
%%\end{deluxetable}
\end{tabular}
\end{center}
Note: Column 1: spectral ID in LAMOST catalog. Column 2: right ascension. Column 3: declination.
Column 4: stellar subclass. Column 5: $V$-band magnitude.
Column 6: extinction in the $V$ band. Column 7: hydrogen column density converted from optical extinction. Column 8: effective temperature.
 Column 9: surface gravity.
Column 10: metallicity.
%Column 15: H$\alpha$ Equivalent Width. Column 16: uncertainty of H$\alpha$ Equivalent Width.
%\end{threeparttable}
%%\end{table}
\end{table}

\begin{table}
\renewcommand{\arraystretch}{1.5}
\tabcolsep=2pt
\scriptsize
\begin{center}
\caption[]{Estimated parameters for the H$\alpha$ and H$\beta$ lines.}
%\label{xxx.tab}
\vspace{0.1cm}
\begin{tabular}{ccccccc}
\hline
obsid& subclass & EW$_{\rm H\alpha}$ & EW$'_{\rm H\alpha}$ & log($f_{\rm H\alpha}/f_{\rm bol}$) & FWHM$_{\rm H_{\beta}, cor}$ & FWHM$_{\rm H_{\alpha}, cor}$ \\
     &          & (\AA) &  (\AA) & & (\AA) & (\AA) \\
(1)&(2)&(3)&(4)&(5)&(6)&(7) \\
\hline
  1603040 &  G9 & -0.44 $\pm$ 0.04 &...&...&...&...\\
  1615104 &  K1 & -0.61 $\pm$ 0.03 &...&...&...&...\\
  1615179 &  F6 & -1.67 $\pm$ 0.08 &...&...&...&...\\
  1615187 &  K3 & -0.45 $\pm$ 0.45 &...&...&...&...\\
  7003074 &  K5 & 1.30 $\pm$ 0.22 & 1.93 & -3.73 $\pm$ 0.05 &...& 1.26 $\pm$ 2.66 \\
  7003243 &  G8 & -0.48 $\pm$ 0.06 &...&...& 3.42 $\pm$ 5.03  &...\\
  7711046 &  K5 & 0.78 $\pm$ 0.13 &...&...&...&...\\
  20001189 &  K3 & -0.88 $\pm$ 0.12 &...&...&...&...\\
  28114180 &  G5 & -1.31 $\pm$ 0.03 &...&...&...&...\\
  28306100 &  G3 & -1.97 $\pm$ 0.04 &...&...& 1.84 $\pm$ 1.53  &...\\
  42815236 &  K0 & -0.72 $\pm$ 0.03 &...&...&...&...\\
  43810009 &  K4 & -0.81 $\pm$ 0.03 &...&...&...&...\\
  51007061 &  G3 & -2.17 $\pm$ 0.05 &...&...& 2.28 $\pm$ 1.11  & 2.29 $\pm$ 1.07 \\
  53111238 &  F6 & -2.56 $\pm$ 0.06 &...&...&...& 3.23 $\pm$ 0.83  \\

\hline
\end{tabular}
\end{center}
Note: Column 1: spectral ID in LAMOST catalog.
Column 2: stellar subclass. Column 3: H$\alpha$ EWs.
Column 4: excess H$\alpha$ EWs. Column 5: H$\alpha$ to bolometric flux ratio.
Column 6: corrected H$\beta$ FWHMs.
Column 7: corrected H$\alpha$ FWHMs.
%\end{threeparttable}
%%\end{table}
\end{table}

\begin{acknowledgements}
This research has made use of data obtained from the 3XMM XMM-Newton serendipitous source catalogue compiled by the 10 institutes of the XMM-Newton Survey Science Centre selected by ESA.
Guoshoujing Telescope (the Large Sky Area Multi-Object Fiber Spectroscopic Telescope LAMOST) is a National Major Scientific Project built by the Chinese Academy of Sciences. Funding for the project has been provided by the National Development and Reform Commission. LAMOST is operated and managed by the National Astronomical Observatories, Chinese Academy of Sciences.
We acknowledge use of the SIMBAD database and the VizieR catalogue access tool, operated at CDS, Strasbourg, France, and of Astropy, a community-developed core Python package for Astronomy (Astropy Collaboration, 2013). We are grateful for support from the National Science Foundation of China (NSFC, Nos. 11273028, 11333004, 11603035, 11603038, and 11503054). RS acknowledges support from a Curtin University Senior Research Fellowship; he is also grateful for support, discussions and hospitality at the Strasbourg Observatory during part of this work.
\end{acknowledgements}


\begin{thebibliography}{}
\bibitem[Ag{\"u}eros et al.(2009)]{2009ApJS..181..444A} Ag{\"u}eros, M.~A., Anderson, S.~F., Covey, K.~R., et al.\ 2009, \apjs, 181, 444
\bibitem[Am{\^o}res et al.(2017)]{2017A&A...602A..67A} Am{\^o}res, E.~B., Robin, A.~C., \& Reyl{\'e}, C.\ 2017, \aap, 602, A67
\bibitem[Auri{\`e}re et al.(2015)]{Auriere2015} Auri{\`e}re, M., Konstantinova-Antova, R., Charbonnel, C., et al.\ 2015, \aap, 574, A90
\bibitem[Baliunas \& Jastrow(1990)]{Baliunas1990} Baliunas, S., \& Jastrow, R.\ 1990, \nat, 348, 520
\bibitem[Brusa et al.(2007)]{2007ApJS..172..353B} Brusa, M., Zamorani, G., Comastri, A., et al.\ 2007, \apjs, 172, 353
%\bibitem[(2014)]{2014} Binney, J., Burnett, B., Kordopatis, G., et al.\ 2014, \mnras, 437, 351
%\bibitem[Burnett \& Binney(2010)]{2010MNRAS.407..339B} Burnett, B., \& Binney, J.\ 2010, \mnras, 407, 339
\bibitem[Cardini \& Cassatella(2007)]{Cardini2007} Cardini, D., \& Cassatella, A.\ 2007, \apj, 666, 393
%\bibitem[Carney et al.(1994)]{1994AJ....107.2240C} Carney, B.~W., Latham, D.~W., Laird, J.~B., \& Aguilar, L.~A.\ 1994, \aj, 107, 2240
%\bibitem[Carroll \& Ostlie(1996)]{1996ima..book.....C} Carroll, B.~W., \& Ostlie, D.~A.\ 1996, Institute for Mathematics and Its Applications
\bibitem[Castelli \& Kurucz(2004)]{ck04} Castelli, F., \& Kurucz, R.~L.\ 2004, arXiv:astro-ph/0405087
\bibitem[Charbonneau(2010)]{Charbonneau2010} Charbonneau, P.\ 2010, Living Reviews in Solar Physics, 7, 3
\bibitem[Ciardi et al.(2011)]{Ciardi2011} Ciardi, D.~R., von Braun, K., Bryden, G., et al.\ 2011, \aj, 141, 108
\bibitem[Cui et al.(2012)]{Cui2012} Cui, X.~Q., Zhao, Y.~H., Chu, Y.Q., et al. \ 2012, Research in Astronomy and Astrophysics, 12, 1197
\bibitem[Deng et al.(2012)]{2012RAA....12..735D} Deng, L.-C., Newberg, H.~J., Liu, C., et al.\ 2012, Research in Astronomy and Astrophysics, 12, 735
%\bibitem[Cutri et al.(2013)]{Cutri2013} Cutri, R.~M., Wright, E.~L., Conrow, T., et al.\ 2013, Explanatory Supplement to the AllWISE Data Release Products, by R.~M.~Cutri et al.
%\bibitem[Du et al.(2012)]{Du2012} Du, B., Luo, A., Zhang, J., Wu, Y., \& Wang, F.\ 2012, \procspie, 8451, 845137

\bibitem[Durney et al.(1993)]{Durney1993} Durney, B.~R., De Young, D.~S., \& Roxburgh, I.~W.\ 1993, \solphys, 145, 207
%\bibitem[Fekel et al.(1986)]{1986ApJS...60..551F} Fekel, F.~C., Moffett, T.~J., \& Henry, G.~W.\ 1986, \apjs, 60, 551
\bibitem[Fekel(1997)]{Fekel1997} Fekel, F.~C.\ 1997, \pasp, 109, 514
\bibitem[Foight et al.(2016)]{Foight2016} Foight, D.~R., G{\"u}ver, T., {\"O}zel, F., \& Slane, P.~O.\ 2016, \apj, 826, 66
\bibitem[Georgakakis et al.(2004)]{2004MNRAS.349..135G} Georgakakis, A., Georgantopoulos, I., Vallb{\'e}, M., et al.\ 2004, \mnras, 349, 135
%\bibitem[Gray(1982)]{1982ApJ...262..682G} Gray, D.~F.\ 1982, \apj, 262, 682
%\bibitem[G{\"u}del et al.(1997)]{1997ApJ...483..947G} G{\"u}del, M., Guinan, E.~F., \& Skinner, S.~L.\ 1997, \apj, 483, 947
\bibitem[G{\"u}del(2004)]{2004A&ARv..12...71G} G{\"u}del, M.\ 2004, \aapr, 12, 71
\bibitem[Harnden et al.(1979)]{Harnden1979} Harnden, F.~R., Jr., Branduardi, G., Gorenstein, P., et al.\ 1979, \apjl, 234, L51
%\bibitem[Hauschildt \& Baron(1999)]{1999JCoAM.109...41H} Hauschildt, P.~H., \& Baron, E.\ 1999, Journal of Computational and Applied Mathematics, 109, 41
\bibitem[Henry et al.(1996)]{Henry1996} Henry, T.~J., Soderblom, D.~R., Donahue, R.~A., \& Baliunas, S.~L.\ 1996, \aj, 111, 439
%\bibitem[Herbig(1985)]{1985ApJ...289..269H} Herbig, G.~H.\ 1985, \apj, 289, 269
\bibitem[Houdebine et al.(2017)]{Houdebine2017} Houdebine, E.~R., Mullan, D.~J., Bercu, B., Paletou, F., \& Gebran, M.\ 2017, \apj, 837, 96
\bibitem[Hornschemeier et al.(2003)]{2003AJ....126..575H} Hornschemeier, A.~E., Bauer, F.~E., Alexander, D.~M., et al.\ 2003, \aj, 126, 575
\bibitem[Jansen et al.(2001)]{Jansen2001} Jansen, F., Lumb, D., Altieri, B., et al. \ 2001, \aap, 365, L1
\bibitem[Jardine \& Unruh(1999)]{Jardine1999} Jardine, M., \& Unruh, Y.~C.\ 1999, \aap, 346, 883
\bibitem[Jeffries et al.(2006)]{Jeffries2006} Jeffries, R.~D., Evans,
P.~A., Pye, J.~P., \& Briggs, K.~R.\ 2006, \mnras, 367, 781
\bibitem[Jenkins et al.(2011)]{Jenkins2011} Jenkins, J.~S., Murgas, F.,
Rojo, P., et al.\ 2011, \aap, 531, A8
\bibitem[Jester et al.(2005)] {Jester2005} Jester, S., Schneider, D.~P., Richards, G.~T., et al.\ 2005, \aj, 130, 873
\bibitem[Judge et al.(2003)]{Judge2003} Judge, P.~G., Solomon, S.~C., \& Ayres, T.~R.\ 2003, \apj, 593, 534
%\bibitem[Johnstone \& G{\"u}del(2015)]{2015A&A...578A.129J} Johnstone, C.~P., \& G{\"u}del, M.\ 2015, \aap, 578, A129
%\bibitem[Koleva et al.(2009)]{Koleva2009} Koleva, M., Prugniel, P., Bouchard, A., \& Wu, Y.\ 2009, \aap, 501, 1269
\bibitem[Katsova \& Livshits(2011)]{Katsova2011} Katsova, M.~M., \& Livshits, M.~A.\ 2011, Astronomy Reports, 55, 1123
\bibitem[Krautter et al.(1999)]{1999A&A...350..743K} Krautter, J., Zickgraf, F.-J., Appenzeller, I., et al.\ 1999, \aap, 350, 743
\bibitem[Lin et al.(2012)]{2012ApJ...756...27L} Lin, D., Webb, N.~A., \& Barret, D.\ 2012, \apj, 756, 27
\bibitem[Luo et al.(2015)]{Luo2015} Luo, A.~L., Zhao, Y.~H., Zhao, G., et al. \ 2015, Research in Astronomy and Astrophysics, 15, 1095
\bibitem[Luo et al.(2008)]{2008ApJS..179...19L} Luo, B., Bauer, F.~E., Brandt, W.~N., et al.\ 2008, \apjs, 179, 19-36
\bibitem[Maccacaro et al.(1988)]{Maccacaro1988} Maccacaro, T., Gioia, I.~M., Wolter, A., Zamorani, G., \& Stocke, J.~T.\ 1988, \apj, 326, 680
\bibitem[Mamajek \& Hillenbrand(2008)]{Mamajek2008} Mamajek, E.~E., \& Hillenbrand, L.~A.\ 2008, \apj, 687, 1264-1293
\bibitem[Mart{\'{\i}}nez-Arn{\'a}iz et al.(2011)]{Martinez-Arnaiz2011}
Mart{\'{\i}}nez-Arn{\'a}iz, R., L{\'o}pez-Santiago, J., Crespo-Chac{\'o}n,
I., \& Montes, D.\ 2011, \mnras, 414, 2629
%\bibitem[Majewski et al.(2011)]{Majewski2011} Majewski, S.~R., Zasowski, G., \& Nidever, D.~L.\ 2011, \apj, 739, 25
%\bibitem[Molenda-{\.Z}akowicz et al.(2013)]{2013MNRAS.434.1422M} Molenda-{\.Z}akowicz, J., Sousa, S.~G., Frasca, A., et al.\ 2013, \mnras, 434, 1422
%\bibitem[Mullan \& Cheng(1994)]{1994ApJ...435..435M} Mullan, D.~J., \& Cheng, Q.~Q.\ 1994, \apj, 435, 435
\bibitem[Noyes et al.(1984)]{1984ApJ...279..763N} Noyes, R.~W., Hartmann, L.~W., Baliunas, S.~L., Duncan, D.~K., \& Vaughan, A.~H.\ 1984, \apj, 279, 763
\bibitem[{\"O}zdarcan \& Dal(2018)]{Ozdarcan2018} {\"O}zdarcan, O., \& Dal, H.~A.\ 2018, arXiv:1801.06087 
\bibitem[Pace(2013)]{Pace2013} Pace, G.\ 2013, \aap, 551, L8
\bibitem[Pallavicini et al.(1981)]{1981ApJ...248..279P} Pallavicini, R., Golub, L., Rosner, R., et al.\ 1981, \apj, 248, 279
\bibitem[Pizzolato et al.(2003)]{Pizzolato2003} Pizzolato, N., Maggio, A., Micela, G., Sciortino, S., \& Ventura, P.\ 2003, \aap, 397, 147
\bibitem[Prosser et al.(1996)]{Prosser1996} Prosser, C.~F., Randich, S., Stauffer, J.~R., Schmitt, J.~H.~M.~M., \& Simon, T.\ 1996, \aj, 112, 1570
\bibitem[Reid et al.(1995)]{Reid1995} Reid, I.~N., Hawley, S.~L., \& Gizis, J.~E.\ 1995, \aj, 110, 1838
\bibitem[Reiners et al.(2014)]{Reiners2014} Reiners, A., Sch{\"u}ssler, M., \& Passegger, V.~M.\ 2014, \apj, 794, 144
\bibitem[Rocha-Pinto \& Maciel(1998)]{1998MNRAS.298..332R} Rocha-Pinto, H.~J., \& Maciel, W.~J.\ 1998, \mnras, 298, 332
\bibitem[Rogel et al.(2006)]{Rogel2006} Rogel, A.~B., Lugger, P.~M., Cohn,
H.~N., et al.\ 2006, \apjs, 163, 160
\bibitem[Rosen et al.(2016)]{Rosen2016} Rosen, S.~R., Webb, N.~A., Watson, M.~G.,et al.\ 2016, \aap, 590, 1
%\bibitem[Schlafly \& Finkbeiner(2011)]{Schlafly2011} Schlafly, E.~F., \& Finkbeiner, D.~P.\ 2011, \apj, 737, 103
%\bibitem[Ryan \& Norris(1991)]{1991AJ....101.1835R} Ryan, S.~G., \& Norris, J.~E.\ 1991, \aj, 101, 1835
\bibitem[Schlafly \& Finkbeiner(2011)]{2011ApJ...737..103S} Schlafly, E.~F., \& Finkbeiner, D.~P.\ 2011, \apj, 737, 103
\bibitem[Schmitt et al.(1990)]{Schmitt1990} Schmitt, J.~H.~M.~M., Collura, A., Sciortino, S., et al.\ 1990, \apj, 365, 704
\bibitem[Schmitt et al.(1995)]{Schmitt1995} Schmitt, J.~H.~M.~M., Fleming, T.~A., \& Giampapa, M.~S.\ 1995, \apj, 450, 392
\bibitem[Schmitt(1997)]{Schmitt1997} Schmitt, J.~H.~M.~M.\ 1997, \aap, 318, 215
\bibitem[Schrijver et al.(1984)]{Schrijver1984} Schrijver, C.~J., Mewe, R., \& Walter, F.~M.\ 1984, \aap, 138, 258
\bibitem[Schr{\"o}der \& Schmitt(2007)]{Schroder2007} Schr{\"o}der, C., \& Schmitt, J.~H.~M.~M.\ 2007, \aap, 475, 677 
%\bibitem[Schrijver(1993)]{Schrijver1993} Schrijver, C.~J.\ 1993, \aap, 269, 446
\bibitem[Skumanich(1972)]{Skumanich1972} Skumanich, A.\ 1972, \apj, 171, 565
\bibitem[Simon \& Drake(1989)]{Simon1989} Simon, T., \& Drake, S.~A.\ 1989, \apj, 346, 303
\bibitem[Simon et al.(1985)]{Simon1985} Simon, T., Herbig, G., \& Boesgaard, A.~M.\ 1985, \apj, 293, 551
\bibitem[Sissa et al.(2016)]{2016A&A...596A..76S} Sissa, E., Gratton, R., Desidera, S., et al.\ 2016, \aap, 596, A76
%\bibitem[Stauffer et al.(1984)]{Stauffer1984} Stauffer, J.~R., Hartmann, L., Soderblom, D.~R., \& Burnham, N.\ 1984, \apj, 280, 202
%\bibitem[Stern et al.(1981)]{Stern1981} Stern, R.~A., Zolcinski, M.~C., Antiochos, S.~K., \& Underwood, J.~H.\ 1981, \apj, 249, 647
\bibitem[Solanki et al.(1997)]{Solanki1997} Solanki, S.~K., Motamen, S., \& Keppens, R.\ 1997, \aap, 325, 1039
\bibitem[Stelzer et al.(2013)]{Stelzer2013} Stelzer, B., Marino, A.,
Micela, G., L{\'o}pez-Santiago, J., \& Liefke, C.\ 2013, \mnras, 431, 2063
\bibitem[Stocke et al.(1983)]{Stocke1983} Stocke, J.~T., Liebert, J., Gioia, I.~M., et al.\ 1983, \apj, 273, 458
\bibitem[Stocke et al.(1991)]{1991ApJS...76..813S} Stocke, J.~T., Morris, S.~L., Gioia, I.~M., et al.\ 1991, \apjs, 76, 813
\bibitem[Strassmeier et al.(1990)]{Strassmeier1990} Strassmeier, K.~G., Fekel, F.~C., Bopp, B.~W., Dempsey, R.~C., \& Henry, G.~W.\ 1990, \apjs, 72, 191
\bibitem[Telleschi et al.(2005)]{Telleschi2005} Telleschi, A., G{\"u}del, M., Briggs, K., et al.\ 2005, \apj, 622, 653
\bibitem[Telleschi et al.(2007)]{Telleschi2007} Telleschi, A., G{\"u}del, M., Briggs, K.~R., Audard, M., \& Palla, F.\ 2007, \aap, 468, 425
\bibitem[Testa et al.(2015)]{Testa2015} Testa, P., Saar, S.~H.,
\& Drake, J.~J.\ 2015, Philosophical Transactions of the Royal Society of London Series A, 373, 20140259
%\bibitem[Vaiana et al.(1981)]{Vaiana1981} Vaiana, G.~S., Cassinelli, J.~P., Fabbiano, G., et al.\ 1981, \apj, 245, 163
\bibitem[Vaiana(1983)]{Vaiana1983} Vaiana, G.~S.\ 1983, Solar and Stellar Magnetic Fields: Origins and Coronal Effects, 102, 165
\bibitem[Vaughan \& Preston(1980)]{1980PASP...92..385V} Vaughan, A.~H., \& Preston, G.~W.\ 1980, \pasp, 92, 385
\bibitem[Vilhu(1984)]{Vilhu1984} Vilhu, O.\ 1984, \aap, 133, 117
\bibitem[Vilhu \& Walter(1987)]{1987ApJ...321..958V} Vilhu, O., \& Walter, F.~M.\ 1987, \apj, 321, 958
\bibitem[Voges et al.(1999)]{1999A&A...349..389V} Voges, W., Aschenbach, B., Boller, T., et al.\ 1999, \aap, 349, 389
\bibitem[Walter \& Bowyer(1981)]{1981ApJ...245..671W} Walter, F.~M., \& Bowyer, S.\ 1981, \apj, 245, 671
%\bibitem[Walter \& Bowyer(1981)]{Walter \& Bowyer1981} Walter, F.~M., \& Bowyer, S.\ 1981, \apj, 245, 671
%\bibitem[Wang et al.(2016)]{Wang2016} Wang, J., Shi, J., Zhao, Y., et al.\ 2016, \mnras, 456, 672
\bibitem[Wegner(2007)]{Wegner2007} Wegner, W.\ 2007, \mnras, 374, 1549
\bibitem[Wright et al.(2004)]{2004ApJS..152..261W} Wright, J.~T., Marcy, G.~W., Butler, R.~P., \& Vogt, S.~S.\ 2004, \apjs, 152, 261
\bibitem[Wright et al.(2011)]{2011ApJ...743...48W} Wright, N.~J., Drake, J.~J., Mamajek, E.~E., \& Henry, G.~W.\ 2011, \apj, 743, 48
\bibitem[Wilson(1963)]{Wilson1963} Wilson, O.~C.\ 1963, \apj, 138, 832
\bibitem[Worthey et al.(1994)]{1994ApJS...94..687W} Worthey, G., Faber, S.~M., Gonzalez, J.~J., \& Burstein, D.\ 1994, \apjs, 94, 687
\bibitem[Wu et al.(2011)]{Wu2011} Wu, Y., Luo, A.-L., Li, H.-N., et al.\ 2011, Research in Astronomy and Astrophysics, 11, 924
\bibitem[Xiang et al.(2017)]{Xiang2017} Xiang, M.-S., Liu, X.-W., Shi, J.-R., et al.\ 2017, \mnras, 464, 3657
\bibitem[Yang et al.(2017)]{2017ApJ...849...36Y} Yang, H., Liu, J., Gao, Q., et al.\ 2017, \apj, 849, 36
%\bibitem[Yong \& Lambert(2003)]{2003PASP..115...22Y} Yong, D., \& Lambert, D.~L.\ 2003, \pasp, 115, 22
\bibitem[Zacharias et al.(2013)]{Zacharias2013} Zacharias, N., Finch, C.~T., Girard, T.~M., et al.\ 2013, \aj, 145, 44
\bibitem[Zhao et al.(2012)]{Zhao2012} Zhao, G., Zhao, Y,~H., Chu, Y.~Q., et al. \ 2012, Research in Astronomy and Astrophysics, 12, 723
\bibitem[Zickgraf et al.(2003)]{2003A&A...406..535Z} Zickgraf, F.-J., Engels, D., Hagen, H.-J., Reimers, D., \& Voges, W.\ 2003, \aap, 406, 535


\end{thebibliography}
\end{document}